\documentclass[a4paper,fleqn]{cas-dc}

\usepackage{xcolor}
\usepackage{natbib}
\usepackage{longtable}
\usepackage{mathtools, cuted}
\renewcommand{\vec}[1]{\mathbf{#1}}

\def\tsc#1{\csdef{#1}{\textsc{\lowercase{#1}}\xspace}}
\tsc{WGM}
\tsc{QE}
\tsc{EP}
\tsc{PMS}
\tsc{BEC}
\tsc{DE}

\begin{document}
\let\WriteBookmarks\relax
\def\floatpagepagefraction{1}
\def\textpagefraction{.001}
\shorttitle{A Robust Model for Flux Density Calculations of Radio Halos: {\tt Halo-FDCA}}
\shortauthors{J.M. Boxelaar et al.}

\title [mode = title]{A Robust Model for Flux Density Calculations of Radio Halos in Galaxy Clusters: {\tt Halo-FDCA}}

\author[1]{J.M. Boxelaar}[type=editor,
                        ]
\ead{boxelaar@strw.leidenuniv.nl}
\ead[url]{ https://github.com/JortBox/Halo-FDCA}
\credit{Conceptualization of this study, Methodology, Software}

\author[1]{R. J. van Weeren}[type=editor,
                        ]
\credit{Conceptualization of this study, Methodology, Software}

\author[1]{A. Botteon}[type=editor,
                        ]
\credit{Conceptualization of this study, Methodology, Software}

\address[1]{Leiden Observatory, Leiden University, Niels Bohrweg 2, 2300 RA Leiden, The Netherlands}

\begin{abstract}
Here we present {\tt Halo-FDCA}, a robust open source {\tt Python} package for modeling and estimating total flux densities of radio (mini) halos in galaxy clusters. Radio halos are extended ($\sim 200-1500$ kpc in size) synchrotron emitting sources found in galaxy clusters that trace the presence of cosmic rays and magnetic fields in the intracluster medium (ICM). These sources are centrally located and have a low surface brightness. Their exact origin is still unknown but they are likely related to cosmic rays being re-accelerated in-situ by merger or sloshing driven ICM turbulence. The presented algorithm combines the numerical power of the Markov Chain Monte Carlo routine and multiple theoretical models to estimate the total radio flux density of a radio halo from a radio image and its associated uncertainty. This method introduces a flexible analytic fitting procedure to replace existing simplistic manual measurements prone to biases and inaccuracies. It allows to easily determine the properties of the emission and is particularly suitable for future studies of large samples of clusters. 
\end{abstract}


\maketitle

\section{Introduction}
\label{sec:introduction}

Galaxy clusters are massive gravitationally bound systems consisting of hundreds to thousands of individual galaxies and have total masses of the order of $\sim10^{14-15}$ M$_{\odot}$. However, most of the baryonic matter in galaxy clusters is in the form of a hot dilute plasma that permeates the cluster's volume and that is called intracluster medium (ICM). The ICM emits thermal bremsstrahlung at X-rays wavelengths \citep{1982ARA&A..20..547F,1986RvMP...58....1S}.

In the radio band, particularly at lower frequencies, extended radio emission is observed in some galaxy clusters \citep[for a recent review see][]{weeren_review}. This diffuse emission is not directly associated with individual cluster member galaxies and indicates the presence of cosmic rays (CR) and magnetic fields in the ICM. This radio synchrotron emission generally has a steep spectrum, with $\alpha \lesssim -1$ ($S\propto\nu^{\alpha}$ where $\alpha$ is the spectral index). Cluster-scale emission in galaxy clusters has been classified into two main categories: radio (mini) halos and relics. The focus of this work are radio (mini) halos which are centrally located and their morphologies approximately follow the X-ray emission from the ICM. They are thought to trace particles accelerated by turbulence in the ICM. Giant radio halos extending over Mpc-scales are thought to trace turbulence induced by energetic cluster merger events \citep{2001MNRAS.320..365B,2001ApJ...557..560P}. In dynamically relaxed clusters, diffuse radio emission covering the central region of the cluster can be found in form of radio mini-halos \citep[e.g.,][]{2002A&A...386..456G,2017ApJ...841...71G}. These sources are smaller in size compared to halos (100s kpc vs Mpc-scale), and they are believed to trace turbulent motions in cluster cores injected by sloshing or by the central AGN \citep[e.g,][]{2008ApJ...675L...9M,2013ApJ...762...78Z}. The precise acceleration mechanisms that operate in the ICM are still a point of debate, although some consensus exists on the global processes  \citep[e.g,][]{icm-turbulence}. 

Radio halos and mini-halos are generally morphologically smooth, although there are several examples of irregular shapes thanks to the advent of increasingly high sensitive observations that unveil substructures within the diffuse emission itself \citep[e.g.,][]{2017MNRAS.469.3872G,2020ApJ...897...93B}. Examples of known and well-studied clusters with central bright and extended radio emission are the Coma cluster, Abell\,2744, and the Perseus cluster \citep[e.g.,][]{1993ApJ...406..399G,1998A&A...331..901S,2011MNRAS.412....2B,2017ApJ...845...81P}. Currently, more than 100 diffuse radio sources in the ICM are known and their number is increasing rapidly due to recent advances of low-frequency radio telescopes and surveys  \citep[e.g.,][]{2009IEEEP..97.1522J,2011PASA...28..215N,LOFAR_survey,duchesne17,2017MNRAS.464.1146H,2017A&A...598A.104S,LoTTS-DR2,2017CSci..113..707G, 2021arXiv210209238D}.

Using large samples of radio halos we can study their statistical properties, such as how the total power and emissivity in the cluster's volume scale with cluster mass. These quantities form an important basis to test theoretical models for the formation of diffuse sources in the ICM. Using a fitted flux density profile, the total flux, surface brightness, and size of a halo can be accurately estimated. Currently, measured values in the literature often adopt a boundary of the radio emission based on a certain contour level determined by the noise of the radio image. This means that the sizes and integrated flux densities do depend on the map noise and grow when deeper observations are available. The main objective in this work is to develop a robust model to fit the flux density profiles of radio halos in galaxy clusters which results in a generic algorithm functioning with a little user input, limiting the biases introduced by manual measurements.%

Our {\tt Halo-FDCA} (Halo-Flux Density CAlculator) builds on the work of \cite{radial_profile,2017MNRAS.470.3465B}. It improves upon previous works as flux density profiles are fitted directly to 2D images instead of first radially averaging them. It also adds additional models that can take into account more complex, asymmetric shapes of the diffuse emission. It uses the Markov Chain Monte Carlo method to provide a robust assessment of the associated uncertainties. The method is generic and can be applied both to radio halos and mini-halos\footnote{For simplicity, in the following we shall use only the term "halo".}. The code is freely available as an open source project, coded in {\tt Python}, on GitHub\footnote{\url{https://github.com/JortBox/Halo-FDCA}}.

This paper is organized as follows in Sect.~\ref{sec:models} we describe the underlying theoretical models for the surface brightness distribution that we fit. This is followed by Sect.~\ref{sec:code} where we describe the implementation of the {\tt Python} code. We show three examples and results of the fitting in Sect.~\ref{sec:results}. We end with a discussion and conclusions in Sects.~\ref{sec:discussion} and~\ref{sec:conclusion} 

\section{Analytic Fitting Models}
\label{sec:models}
There are already existing methods to measure and estimate the flux density of a radio halo. The most commonly used method is to manually measure the flux in regions around the radio halo by following a certain radio surface brightness  (``contour'') level. Often, values around 2$\sigma_{\rm{rms}}$ or $3\sigma_{\rm{rms}}$ are used, where $\sigma_{\rm{rms}}$ is the root mean square noise in the radio image. The image noise used here is calculated using the \texttt{findrms} function taken from the LOFAR \texttt{ddf-pipeline}\footnote{\url{https://github.com/mhardcastle/ddf-pipeline}}.

A second method is to fit a certain functional form to the radio emission to determine its flux density. \cite{radial_profile} did this by first azimuthally averaging the surface brightness in concentric spherical annuli and then fitting a spherically symmetric exponential model to the one-dimensi\-onal radial surface brightness profile of the form
\begin{equation}
	I(r)=I_0e^{-r/r_e}
	\label{eq:murgia_exp}
\end{equation}
where $r_e$ is a characteristic $e$-folding distance to the halo centre and $I_0$ is the central surface brightness. %
Since the introduction of this model, it has been used a number of times for flux calculations of halos \citep[e.g.][]{orru-2007, vacca-2014}. There are however examples of halos that do not follow this idealized circular distribution: some may have elongated morphologies, e.g. the Toothbrush cluster \citep{2012A&A...546A.124V,2018ApJ...852...65R} and El Gordo \citep{2014ApJ...786...49L,2016MNRAS.463.1534B}, while others show even asymmetric distributions, e.g. \\* MACS\,J0717.5+3745 \citep{2017ApJ...835..197V} and Abell\,2744 \citep{2017ApJ...845...81P}. It has been suggested to compose a more general and robust model to fit profiles to a wide variety of radio halos \citep{2017MNRAS.470.3465B}. 

The main advantage of fitting models to the radio halo emission is that the integrated flux density should not be directly dependent anymore on the depth of the observations and choice of ``contour'' level. %
Measuring the flux density within a region delimited by a certain $\sigma_{\rm{rms}}$ contour level generally leads to larger values if this methodology is applied to another image of the same cluster with a lower noise level (e.g. a deeper follow-up observation), as the halo may appear larger. Vice versa, a high noise level image of the same halo would lead to a smaller measured radius and lower flux density estimate. Using models, we obtain a proper measure of the size of the halo that can be defined and reproduced. The sizes of halos have been commonly measured using a $\sigma_{\rm{rms}}$ contour level, leading to different sizes depending on the noise map.

To construct the analytic models, we start by defining the unconvolved model denoted as $I(\vec{r})$, with $\vec{r} = (x,y)^T$ Cartesian image coordinate vector. The convolved model that is actually fitted to a data image with a certain synthesized beam shape $B_{\phi}$ is given by 
\begin{equation}
	\mathcal{I}(\vec{r}) =  I(\vec{r}) * B_{\phi}. 
\end{equation}
For the beam shape, we take
\begin{equation}
	B_{\phi}(x,y)=\frac{b_xb_y}{4\pi\ln2}
	\exp\left[-4\ln2\left(\frac{X_{\phi}^2}{b^2_x}+\frac{Y_{\phi}^2}{b^2_y}\right)\right]
	\label{eq:beamshape}
\end{equation}
as defined in \citet{2016era..book.....C}. In this definition, $b_x$ denotes the Full-Width at Half-Maximum (FWHM) of the beam major axis, $b_y$ the FWHM of the beam minor axis and $\phi$ the counterclockwise shape rotation. This same rotation transforms a traditional coordinate system $(x,y)$ to
\begin{align}
 \begin{pmatrix}
  X_{\phi} \\
  Y_{\phi}  
 \end{pmatrix} \equiv\vec{R}_{\phi}\vec{r}= 
 \begin{pmatrix}
  \cos{\phi} & \sin{\phi} \\
  -\sin{\phi} & \cos{\phi}
 \end{pmatrix} 
 \begin{pmatrix}
  x \\
  y  
 \end{pmatrix}.
 \label{eq:rotation-matrix}
\end{align}
Furthermore, $\text{FWHM}=\sigma\sqrt{8\ln2}$, with $\sigma$ the Gaussian standard deviation. These are the same quantities as commonly defined in radio observations.

In general, we can define an exponential model to be 
\begin{equation}
	I(\vec{r}) = I_0e^{-G(\vec{r})}
\label{eq:exponential}
\end{equation}
where $I(\vec{r})$ is a surface brightness and $G(\vec{r})$ is the function that takes different forms depending on the complexity of the model chosen by the user in {\tt Halo-FDCA}. We will now introduce the three main models implemented in the program.

\paragraph{Circular.} In the case of the circular model, this function will take the simple form $G(\vec{r})=|\vec{r}|/r_e$, where $|\vec{r}|$ denotes the length of positional vector $\vec{r}$. Coordinate offsets for the different models are implemented by simply replacing $\vec{r}$ by $\vec{r}-\vec{r_0}$ with $\vec{r_0}$ the halo centre (or halo location; $\vec{r_0}=(x_0,y_0)^T$). This form then depicts a 4 parameter model with parameter set $\theta=\{I_0,x_0,y_0,r_e\}$.

Increasing complexity for this model is done analogously to how a two-dimensional circular Gaussian can be generalized. Both a standard two-dimensional Gaussian and $I(\vec{r})$ are Gaussian types with different form factor $k$. A circular Gaussian is defined as
\begin{equation}
 	A\exp \left[-\left({\frac {x^{2}+y^{2}}{2\sigma^{2}}}\right)^{0.5+k}\right]
\end{equation}  
with $k=0.5$ and $A$ the amplitude. In our exponential case, $k=0$ and $\sqrt{2}\sigma$ would be replaced with $r_e$. In this discussion, we explicitly work with a general form that includes the $k$ form factor. %
The presented models and algorithm allow for fitting the $k$ factor as an extra parameter to be able to acquire an insight into the form factor of radio halos. Based on the needs of the user, $k$ can be adopted as an optional parameter. By default, $k=0$ and will not be treated as a fitting parameter by {\tt Halo-FDCA}. Including $k$, the circular model becomes

\begin{equation}
	G(\vec{r})=\left(\frac{|\vec{r}|^2}{r_e^2}\right)^{0.5+k}
	\label{4D-model}
\end{equation}
 
\begin{figure*}[h!]
\centering
   \includegraphics[width=0.50\textwidth]{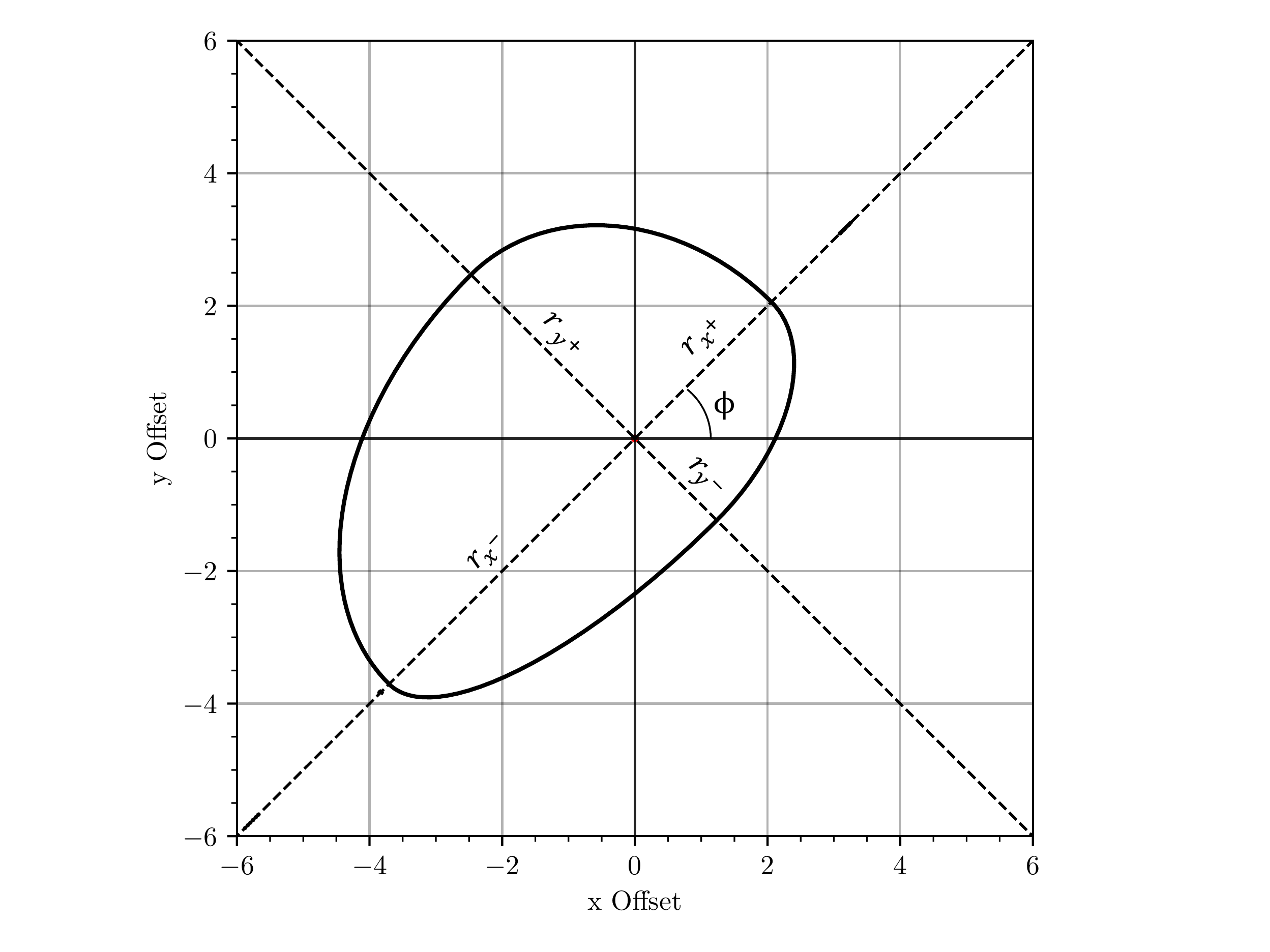}
	\includegraphics[width=0.49\textwidth]{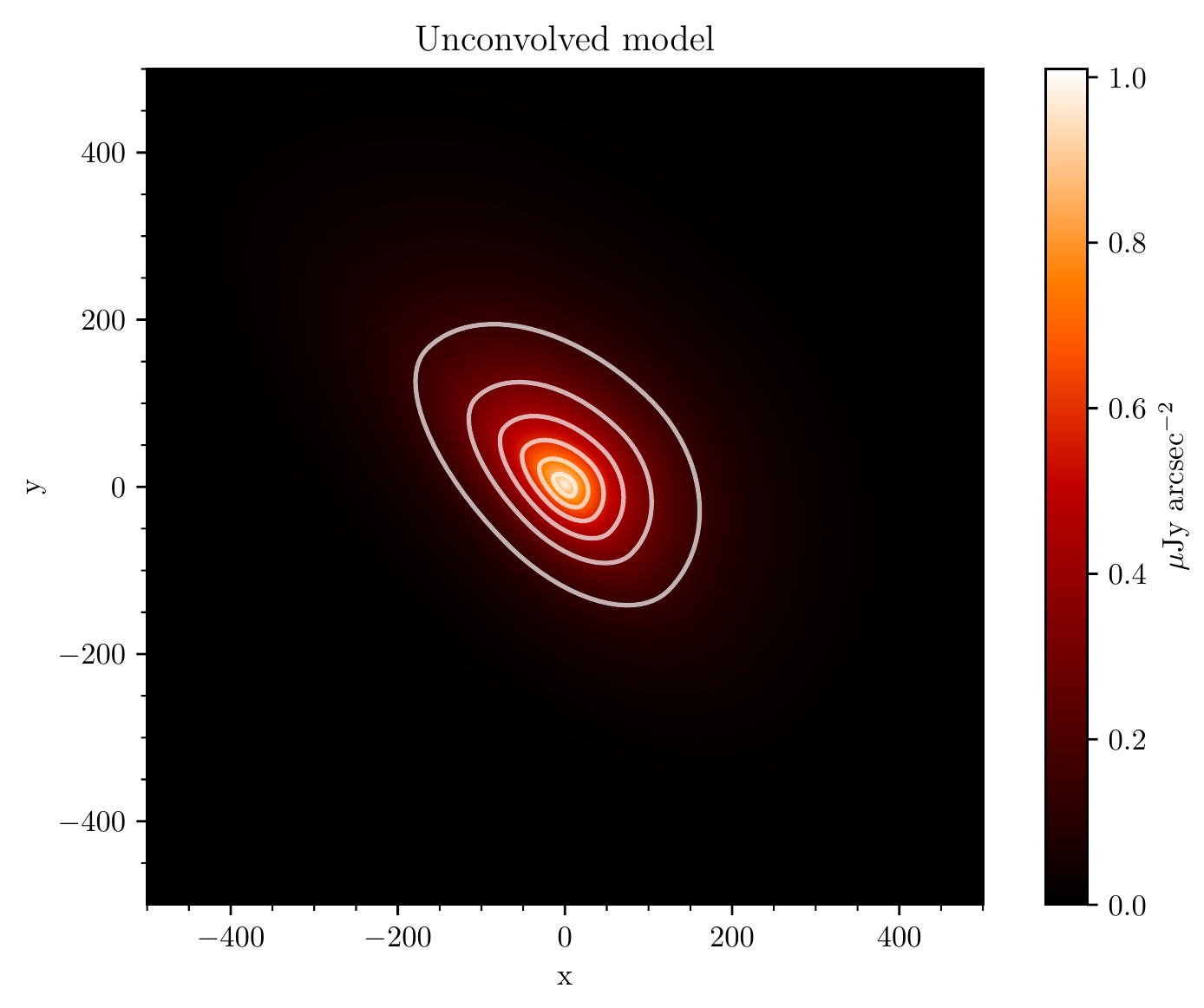}
		\caption{\footnotesize \textit{Left panel}: Illustration of the contour of the skewed model. The main axes of the skewed shape are indicated with dotted lines. Meanings of parameters $r_{x^+}$, $r_{x^-}$, $r_{y^+}$, $r_{y^-}$, and $\phi$ are also indicated.
		\textit{Right panel}: This is an example of what a skewed, 8 parameter exponential profile could look like for $k=0$. In this particular example, the parameter set is $\hat{\theta}=\{1, 0, 0, 75, 50, 120, 90, \pi/4\}$. where $I_0$ is in mJy arcsec$^{-2}$ and all the radii are in pixel units. This image was created using equation \eqref{eq:8D-model}.}
	\label{fig:8D-model}
\end{figure*}

\paragraph{Ellipsoid.} So far the model discussed here (with $k=0$) is the same as in \cite{radial_profile}. From here, a few generalizations are introduced. The model can be extended by separating the two principal axes into two terms and by allowing for rotation $\phi$ with respect to the coordinate system.  Such a model is defined via 

\begin{equation}
	G(\vec{r})=\left[\left(\frac{X_\phi(\vec{r})}{r_x}\right)^2 +\left(\frac{Y_\phi(\vec{r})}{r_y}\right)^2\right]^{0.5+k} ,
\label{6D-model}
\end{equation} 

The set of parameters for this six dimensional model is $\theta=\{I_0,x_0,y_0,r_x,r_y,\phi\}$. 

\paragraph{Skewed.} The next generalization is the construction of a skewed model that allows for an off-center maximum of the intensity distribution. To construct a skewed model, the euclidean plane is divided into its four primary quadrants. The skewed model is then made up of functions defined on a single quadrant only, this is visualized in 
Fig.~\ref{fig:8D-model}. In theory, one must be cautious and think carefully about exactly where the function is defined and how these functions connect to each other at the edge of a quadrant (the $x=0$ and $y=0$ lines). 

The division is made because, in this way, every quadrant can have its own exponential function with two $e$-folding radii at the edge. The first quadrant for example, has radii for the positive $x$ and $y$-axis. To make the overall distribution 'smooth', the functions of each quadrant have equal radii where their edges coincide. Furthermore, all 4 sub-functions are required to have an equal $I_0$. 

We define new parameters $r_{x^+}$, which is defined as the radius in the positive $x$ direction and $r_{x^-}$, which is defined as the radius in the negative $x$ direction. In a similar fashion, $r_{y^+}$ and $r_{y^-}$ are defined. Equation~\eqref{eq:8D-model} provides the mathematical representation of the function. It can also be defined differently but this form simplifies the implementation in the algorithm.

Contrary to previous models, this one is piece-wise continuous and consists of four functions only defined on a specific quadrant of the euclidean plane. This makes integrating this function somewhat more complex. The eight parameter $G$ function is given by

\begin{equation}
  G(\vec{r})^{1/(0.5+k)}=\begin{cases}
    \left(\dfrac{X_{\phi}}{r_{x^{+}}}\right)^2+\left(\dfrac{Y_{\phi}}{r_{y^{+}}}\right)^2, & \text{if $X$, $Y\geq0$}.\\
    
    \left(\dfrac{X_{\phi}}{r_{x^{-}}}\right)^2+\left(\dfrac{Y_{\phi}}{r_{y^{+}}}\right)^2, & \text{if $X\leq0$, $Y>0$}.\\
    
    \left(\dfrac{X_{\phi}}{r_{x^{-}}}\right)^2+\left(\dfrac{Y_{\phi}}{r_{y^{-}}}\right)^2, & \text{if $X$, $Y<0$}.\\
    
    \left(\dfrac{X_{\phi}}{r_{x^{+}}}\right)^2+\left(\dfrac{Y_{\phi}}{r_{y^{-}}}\right)^2, & \text{if $X>0$, $Y\leq0$}.
  \end{cases}
\label{eq:8D-model}
\end{equation}

$G(\vec{r})$ produces the most general exponential profile. $X_{\phi}(\vec{r})$ and $Y_{\phi}(\vec{r})$ account for the offset and rotation of the elliptical shape. The four piece-wise functions are used to introduce skew to the function. The fitting Parameter set for this model is $\theta = \{I_0, \;x_0, \;y_0, \;r_{x^{+}}, \;r_{y^{+}}, \;r_{x^{-}}, \;r_{y^{-}}, \;\phi\}$.

The most general form of the exponential (Equation~\eqref{eq:8D-model}) reduces to Equation~\eqref{6D-model} by setting $r_{x^+}=r_{x^-}\Rightarrow r_x$ and $r_{y^+}=r_{y^-}\Rightarrow r_y$. This can further be simplified to a 5 parameter model by setting $\phi=0$. Equation~\eqref{4D-model} will be found when $r_{x^+}=r_{x^-}=r_{y^+}=r_{y^-}\Rightarrow r_e$ and $\phi=0$.

\subsection{Profile Integrals}
The total halo flux density can be calculated by integrating the analytic functions from the former section. The total flux of a radio halo $S_{\nu}$ at frequency $\nu$  is given by
\begin{equation}
	S_{\nu} = \int\int I(\vec{r})\;dx\;dy.
	\label{eq:integral}
\end{equation}

There might be cases where integrating up to infinity is not desired. In that case it is possible to integrate up to an arbitrary distance $d$. It is important to note that the flux density is integrated from the unconvolved model rather than from the convolved one. This can be done because the integral over both functions are equal to each other thanks to the fact that the beam is normalized. An integral $s_i$ over one quadrant with $e$-folding radii $a$ and $b$ can be expressed in terms of the lower incomplete gamma function $\gamma(x,y)$:

\begin{align*}
    s_i = \frac{\pi r_xr_y}{m} \int\limits_0^{d^{2m}} u^{1/m-1}e^{-x}\;dx=\frac{\pi r_xr_y}{m}\gamma(1/m,d^{2m})
\end{align*}
where for simplicity, $0.5+k=m$. To ensure convergence, we set $m>0$.   In the above expression, $d$ denotes the radius of integration in units of e-folding radius $r_e$. When integrating up to infinity, however, the limit $d\rightarrow\infty$ is taken such that 
 \begin{align*}
     \lim_{d\rightarrow\infty}\gamma(1/m,d^{2m}) = \Gamma(1/m)
 \end{align*}
where $\Gamma$ is the complete Gamma function, whose definition can be deduced from the equation above. With the 8 parameter skewed $G$ function, $S_{\nu}$ takes the form:

\begin{equation}
	S_{\nu} = \frac{I_0\pi}{4m}\Gamma\left(\frac{1}{m}\right)[r_{x^+}r_{y^+}+r_{x^-}r_{y^+}+r_{x^+}r_{y^-}+r_{x^-}r_{y^-}]
\end{equation}
which depends on the $k$ form factor. For the conventional $k=0$ this reduces to the simpler form  
\begin{equation}
	S_{\nu} = \frac{I_0\pi}{2}(r_{x^+}r_{y^+}+r_{x^-}r_{y^+}+r_{x^+}r_{y^-}+r_{x^-}r_{y^-}).
\end{equation}
The extended derivation of the integral over Equation~\eqref{eq:exponential} with Equation~\eqref{eq:8D-model} can be found in Appendix \ref{app:exponential_functions}. From there, simplifying that expression will give us the integrals for the simpler functions. 

Conventionally, analytic models are not integrated up to infinity, but adopt a radius of three times the found $e$-folding radius \cite[see][]{radial_profile}. In that case, one would take $d=3$ and thus $S_{\nu}^{(3)} \simeq 0.8S_{\nu}^{(\infty)}$ for $k=0$ (this can be checked by plugging in $d=3$ in footnote \ref{footnote:appendix} in Appendix \ref{app:exponential_functions}). This is independent of found parameters, which means the flux as integrated to $3r_e$ is 80\% of the flux with integration to infinity for $k=0$ for any obtained set of parameters.

From the total flux of a certain model, the radio power $P_{\nu}$ can be calculated. From $S_{\nu}$, the power is given by

\begin{equation}
	P_{\nu} = \dfrac{4\pi D_L^2}{(1+z)^{1+\alpha}} S_{\nu}
\end{equation}
where $D_L$ is the luminosity distance. To calculate values in physical units (e.g. kpc), a flat $\Lambda$CDM cosmology with $\Omega_{\Lambda} = 0.7, \Omega_m = 0.3, H_0 = 70$ km s$^{-1}$ Mpc$^{-1}$ is assumed in {\tt Halo-FDCA}. A conversion from an observed frequency $\nu_{\mathrm{obs}}$ to the desired frequency can be made using $S_{\nu}\propto \nu^\alpha$.
The code has a built-in function to retrieve the radio power.

\section{{\tt Halo-FDCA}: Code Implementation}
\label{sec:code}

\begin{figure*}[h]
\centering
	\includegraphics[width=0.8\linewidth]{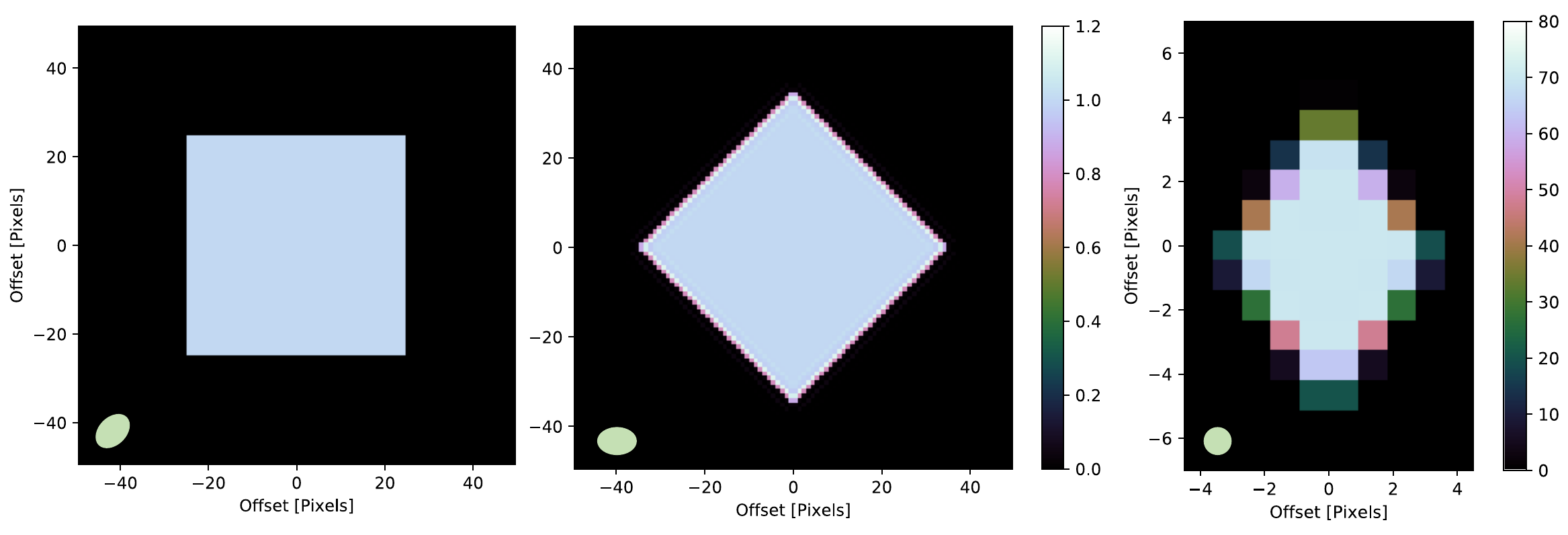}
		\caption{\footnotesize Illustration of the regridding of data to a state where pixel areas approximately equal the image beam area. The left and middle panel have equal colour scales and are in a single, arbitrary unit. The original image depicts a square with constant amplitude 1. The last panel shows a central intensity of 70. First, the image is rotated according to the beam's positional angle (in green in the bottom left of every panel). The new pixel size and aspect ratio are then derived from the beam's minor and major FWHM axis. The image is then regridded to the new pixel size. The resulting image is used by the MCMC.}
	\label{fig:regridding}
\end{figure*}

\subsection{Input data for the fitting}
To properly separate radio halo emission from contaminating sources sufficient spatial resolution is required. The images used for the radio halo profile fitting are expected to be free from other radio sources. Two approaches can be taken to achieve this. This first approach is to subtract all discrete sources from the visibility data by building a model of the contaminating sources in the field using appropriate uv-cuts. This approach is commonly used and can be successful as long as there are no extended radio sources embedded within the radio halo emission. 

A second approach is to mask all regions affected by radio sources other than the radio halo. The code is capable of fitting halos where significant portions are left out by a mask. The algorithm simply extrapolates the model over the regions where data is censored. In practice, the two methods must be combined so that the compact radio sources are subtracted and the extended sources are masked. {\tt Halo-FDCA} accepts optional DS9 \citep{2003ASPC..295..489J} region files to mask areas where the subtraction of radio sources has been inaccurate, or where other types of diffuse emission or residual calibration artifacts are present. To investigate further the influence of masking parts of the image, we run the code on a simple radio halo applying different masks. This is discussed in Section~\ref{sec:mask_results}

Finally, the radio halo emission in the input images is required to be deconvolved and restored, otherwise, the flux measurements can be highly overestimated in particular for arrays with dense inner uv-coverage and sparser coverage in the outer parts of the uv-plane (such as the GMRT and LOFAR). To achieve deep deconvolution, multi-scale clean \citep{2008ISTSP...2..793C,2017MNRAS.471..301O} is advised. The description of pipeline and the up-to-date list of required and optional parameters can be found on the \texttt{Halo-FDCA} GitHub page ( \url{https://github.com/JortBox/Halo-FDCA}).

\subsection{Markov Chain Monte Carlo Algorithm}
\label{sec:MCMC}
For fitting a flux density profile, a Markov Chain Monte Carlo (hereafter MCMC) simulation is used within {\tt Halo-FDCA}. This was done in \texttt{Python 3+} using the \texttt{emcee} module \citep{emcee}, which is based on the theory described in \citet{goodman}. This package is a tool for probabilistic data analysis and model fitting using Bayesian inference to, in this case, find model parameters that maximize a likelihood function $\mathcal{L}$. We adopt the following definition for the log-likelihood function
\begin{align}
	\ln{\mathcal{L}_n(\theta)}=-n\ln{(\sqrt{2\pi}\sigma_{\text{rms}})}-\sum_{i=1}^{n}\dfrac{\left(V(\vec{r}_i)- \mathcal{I}(\vec{r}_i;\theta)\right)^2}{2\sigma_{\text{rms}}^2}.
\label{likelihood}
\end{align}
Here, $V(\vec{r})$ is the  image containing the radio halo and $n$ denotes the total number of pixels.

Bayesian inference is a statistical approach to parametric fitting. It is based on the \emph{belief} of an event happening. Finding best-fit parameters for any parametric model in Bayesian inference can be done by solving Bayes' theorem
\begin{align}
	\mathbb{P}(\Theta=\theta|X=x)&=\frac{\mathbb{P}(X=x|\Theta=\theta)\mathbb{P}(\Theta=\theta)}{\mathbb{P}(X=x)}
\end{align}
with $X, \Theta$ random variables. Furthermore $\mathbb{P}(\Theta=\theta)$ is called the prior and $\mathbb{P}(\Theta=\theta|X=x)$ the posterior. In other words, the equation reads that given the data $x$, the probability that a parameter set $\theta$ describes the model is equal to the probability of the data given the model, times the probability of that set of parameters can exist, divided by the probability that the data is observed. Given a model $f$ and independent and identically distributed data $x_i$, the best fit parameters $\theta$ are found by maximizing 
\begin{align}
	f(\theta|x_1,\dots,x_n)&\propto f(x_1,\dots,x_n|\theta)f(\theta)\\
	&=\prod_{i=1}^n f(x_i|\theta)f(\theta)=\mathcal{L}_n(\theta)f(\theta).
\end{align}

In this last equation, the term corresponding to $\mathbb{P}(X=x)$ was dropped since it is a constant independent of the model parameters, thus not affecting the location of the maximum in equation~\eqref{likelihood}.

\begin{figure*}[h]
	\includegraphics[width=0.8\linewidth]{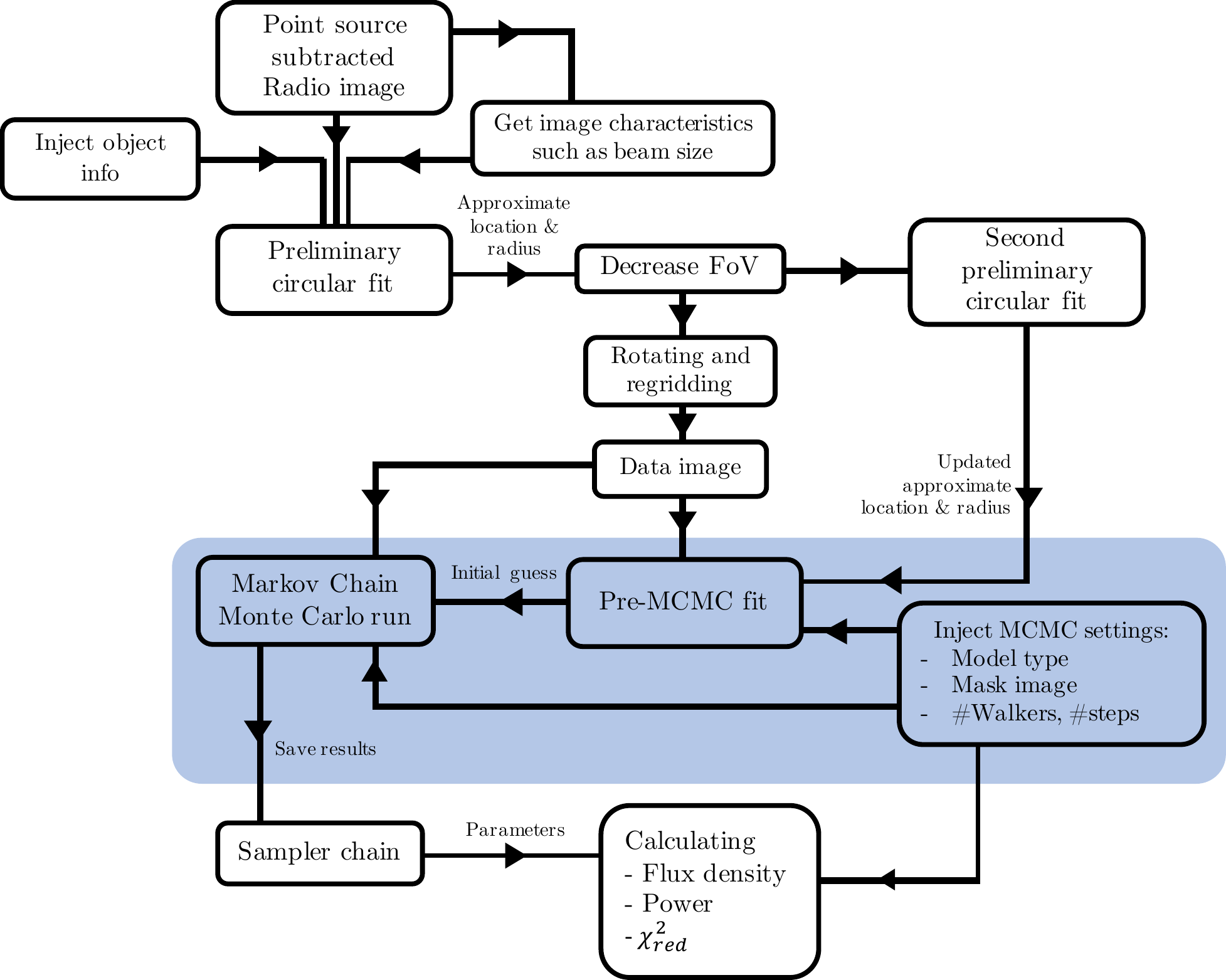}
		\caption{\footnotesize This flowchart gives a schematic representation of the key processes in the code. The boxes within the larger blue cluster represent processes directly related to the MCMC run. The algorithm takes compact source subtracted radio images and, in principle, returns the MCMC sampler chain which is the distribution of best fit parameters. With this chain and the information of the MCMC settings, all the relevant results  such as radio flux and power can be calculated.}
	\label{fig:flowchart}
\end{figure*}

This definition for the likelihood function is derived from assuming that the data is described by $V(\vec{r}_i) = [I(\vec{r}_i ;\hat{\theta})+\kappa(\vec{r}_i)]*B_{\phi}$, where $\kappa$ is the underlying noise map:
\begin{equation}
	\kappa(\vec{r}_i)\overset{\text{iid}}{\sim}\mathcal{N}(0,\sigma_{\text{rms}}^2).  
	\label{eq:noise}
\end{equation}
which is normally independent and identically distributed (iid) around zero. This approximation makes the definition of the likelihood function relatively easy. How close the actual noise distribution is to a normal distribution depends on the image quality. Equation~\eqref{eq:noise} implies that independent data points are required in order to fit models than the data. Generally, pixels values in radio images are correlated and cannot be considered independent values. To assure the independence of data points, data images are rotated and regridded based on the image beam such that pixel areas are equal to the beam area. In practice, new pixel sizes are calculated through

\begin{align*}
    x_{\text{pixel scale}}=b_x\sqrt{\frac{\pi}{4\ln{2}}}\;, && y_{\text{pixel scale}}=b_y\sqrt{\frac{\pi}{4\ln{2}}}.
\end{align*}
In this procedure, the total flux density is preserved. These procedures are included in the flowchart in the next section (Fig. \ref{fig:flowchart}) The total process of rotating and regridding images is visualized in Fig. \ref{fig:regridding}. Regridding inevitably results in using images with different aspect ratios than the original data due to the often elliptical beam shape. An extreme case of this is for the example reported in Section~\ref{sec:phoenix} for the Phoenix cluster, where the beam shape is very elongated.

The \texttt{emcee} code works by exploring the parameter space efficiently with \emph{walkers}. These walkers jump from value to value using a certain move algorithm. By default, the \emph{Stretch Move} is used by \texttt{emcee}. An MCMC run samples the parameter space and results in a chain with all the sample information in it. From these chains, one can determine the resulting maximum likelihood parameters and their uncertainties. The technical and theoretical descriptions of \texttt{emcee} are outside the scope of this paper, it can be found in \cite{emcee} and \cite{goodman}.

\subsubsection{Model Selection and Parameter Dependence}
Every model will in practice result in different flux density measurements. The question then arises which model is the best one to describe the data. In this case, a useful indicator is the reduced $\chi^2$ ($\chi^2_{\mathrm{red}}$). 

The $\chi^2_{\mathrm{red}}$ is a frequently used value that provides insight into the `goodness' of a fit, with 1 typically indicating a good fit. If $\chi^2_{\mathrm{red}}<1$, the model is over-fitting and if $\chi^2_{\mathrm{red}}>1$, the model is under-fitting. In this case $\chi^2_{\mathrm{red}}$ is defined as 

\begin{equation}
	\chi^2_{\text{red}}=\frac{1}{N}\sum_{i=1}^n \frac{(V(\vec{r}_i)-\mathcal{I}(\vec{r}_i;\theta))^2}{\sigma^2_{\text{rms}}}.
\end{equation}

\subsubsection{Image cropping and initial parameter guesses}
In this section, we provide an overview of the main processes of {\tt Halo-FDCA}, such as pre-fitting methods and probability priors. This section is summarised in the flowchart in Fig. \ref{fig:flowchart}. 

Before running the MCMC algorithm, the data is slightly processed as preparation with the aim of decreasing run time. Because a convolved version of the analytic model is fitted to the image, one likelihood function evaluation involves two two-dimensional Fourier transforms. Models also have to be regridded at every likelihood evaluation.

Since the halo often only covers a small part of the radio image, one way to shorten run times and increase the chance of convergence consists in simply cropping the image. %
A second advantage of using cropped images is that some contaminating sources present in images are removed and do not have to be masked manually. This step is shown in Figure~\ref{fig:flowchart} as "Decrease FoV".

When an image is given in input to {\tt Halo-FDCA}, the program does not immediately know where the halo is located. Together with the location as provided by the user, a preliminary fit can be performed to approximate the location and size of the halo on the image. This fit is performed with the simplest model (Eq. \eqref{eq:murgia_exp}) without convolution. The sole purpose of this first fit is to get an idea of the approximate size, surface brightness, and location of the halo. Based on these findings, the image is cropped (when possible) to a size of 8 times the approximated $r_e$ radius, which is enough to prevent any loss of relevant data. The radius and location are fitted again in this cropped image to set up an initial parameter guess for the MCMC run.

The \emph{Pre-MCMC Fit} is the first time the actual convolved model is fitted to the input data. It takes the simple initial guesses provided by the preliminary fit. For the skewed model, this means equal $e$-folding radii in all directions and a rotation of zero. The results of this pre-MCMC fit are taken as the fairly accurate initial guesses for the Markov-chain Monte Carlo algorithm. The pre-MCMC fit is also the stage at which the \emph{mask} (a DS9 region file) is taken into account. With such a mask, specific regions in the data can be ignored by the algorithm. 

\subsubsection{Priors}
Bayesian inference allows us to work with \emph{priors}, which can be defined as $\mathbb{P}[\mathcal{I}(\vec{r};\hat{\theta})]$, the chance of the model for a certain set of parameters. Using the Indicator function $\mathbb{1}$ from statistics, we can define the prior to be 
\begin{equation}
	\mathbb{P}[\mathcal{I}(\vec{r};\hat{\theta})]=\prod_{i=1}^n \mathbb{1}_i
\end{equation}
where $\mathbb{1}_i$'s are given by
\begin{align*}
	&\mathbb{1}_1[\;\hat{I}_0 >0\;], \\
	&\mathbb{1}_2[\;0<\hat{x}_0<x_{\text{max}}\;], \\
	&\mathbb{1}_3[\;0<\hat{y}_0<y_{\text{max}}\;],\\
	&\mathbb{1}_4[\;\hat{r}_{x^+},\;\hat{r}_{y^+},\;\hat{r}_{x^-},\;\hat{r}_{y^-} \geq 0\;],\\
	&\mathbb{1}_5[\;(\hat{r}_{x^+}+\hat{r}_{x^-})<x_{\text{max}}\;,(\hat{r}_{y^+}+\hat{r}_{y^-}) < y_{\text{max}}\;],\\
	&\mathbb{1}_6[\;0\leq\hat{\phi} <\pi \;].
\end{align*}
These are the basic constraints on the parameters. $x_{\text{max}}$ and $y_{\text{max}}$ denote the length of the working image in both dimensions. The last inequality could also be set to $0\leq\phi <2\pi$ but we stick to the first because parameter solution sets are not unique when $0\leq\phi <2\pi$ instead of $0\leq\phi <\pi$. In practice though, an angle prior $-\pi/4\leq\hat{\phi} <5\pi/4 $ is chosen to prevent MCMC confusion around $\phi = 0 = 2\pi$. Inequality five makes sure that what is mathematically defined as the major (x) and minor (y) axis remains associated with that axis.
An extra prior requires $r_x>r_y$ such that the major axis is always on the x-axis. This requirement limits the parameter freedom and allows for a unique parameter solution.

One can also choose to adopt a prior on the form factor $k$ when it is used as an extra parameter. It must be greater than $-1/2$ to ensure a converging flux density integral (by default $k=0$). There is no upper boundary on this extra parameter. %

\subsection{Output}
\label{sec:output}
An MCMC simulation deploys walkers to explore the possible parameter space. After a fitting run, the walker chain (all values explored by a walker) is used to estimate the parameters. The way these walkers behaved and moved through the parameter space is the primary indication of whether the algorithm converged and detected a radio halo. Two types of figures are in that case important; the walker and corner plots. An example of each of these plots is shown in Figure~\ref{fig:walker} and \ref{fig:corner}, respectively. 
 
The walker plot shows in what way the parameter space is sampled. The figure shows a fit for the elliptical model. A total of 200 walkers explore a certain set of parameters 1200 times. The walker plot can be used to visualize the burn-in time: the number of steps that need to be taken before the algorithm settles on a value. The walker plot then also indicates whether walkers settle on one specific value or rather jump between multiple values that have similar likelihood values. Both example figures show that the fitting procedure converged and settled on a set of optimal parameters with a certain uncertainty. %

 \begin{figure*}[t]
	\includegraphics[width=1.7\columnwidth]{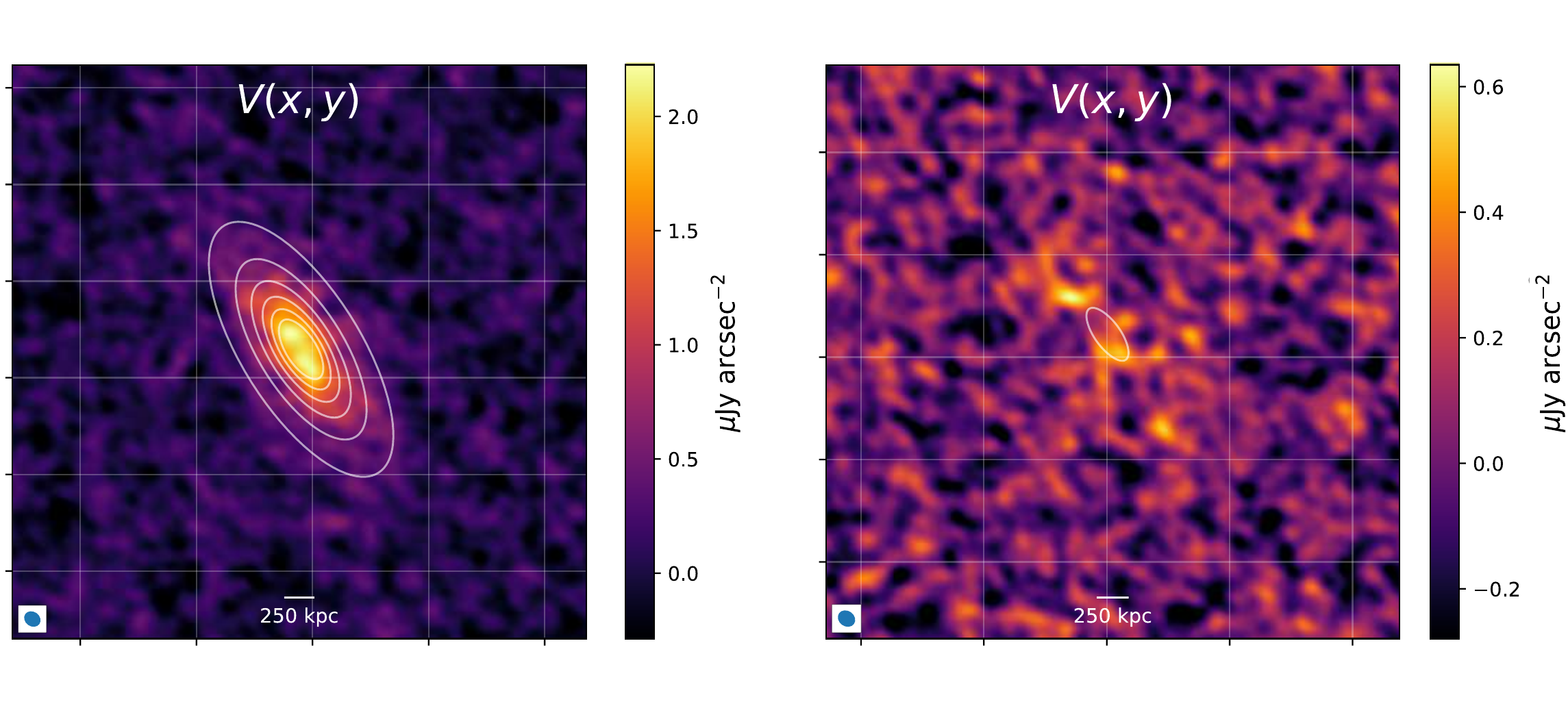}
	
		\caption{\footnotesize Two examples of an artificial halo with different total flux densities. Both panels show the beam shape in the lower left corner and the physical scale. Surface brightness is indicated in the colour bars. \emph{Right:} artificial halo with $S_{\nu}=11.75$ mJy,  this halo is so faint that it has become almost invisible to the  eye. The algorithm though did find a flux density value that is in accordance with what is expected based on the left panel of Fig.~\ref{fig:artificial-testing} \emph{Left:} identical artificial halo with $S_{\nu}=100$ mJy, This is the brightest halo realization. Here the halo is clearly visible. Contours show the models' $[1,2,3,...]\times\sigma_{\text{rms}}$ levels.}
	\label{fig:artificial_example}
\end{figure*}

 \begin{figure*}[th]
	\includegraphics[width=0.68\columnwidth,trim={0.1cm 0.2cm 1.cm 0cm},clip]{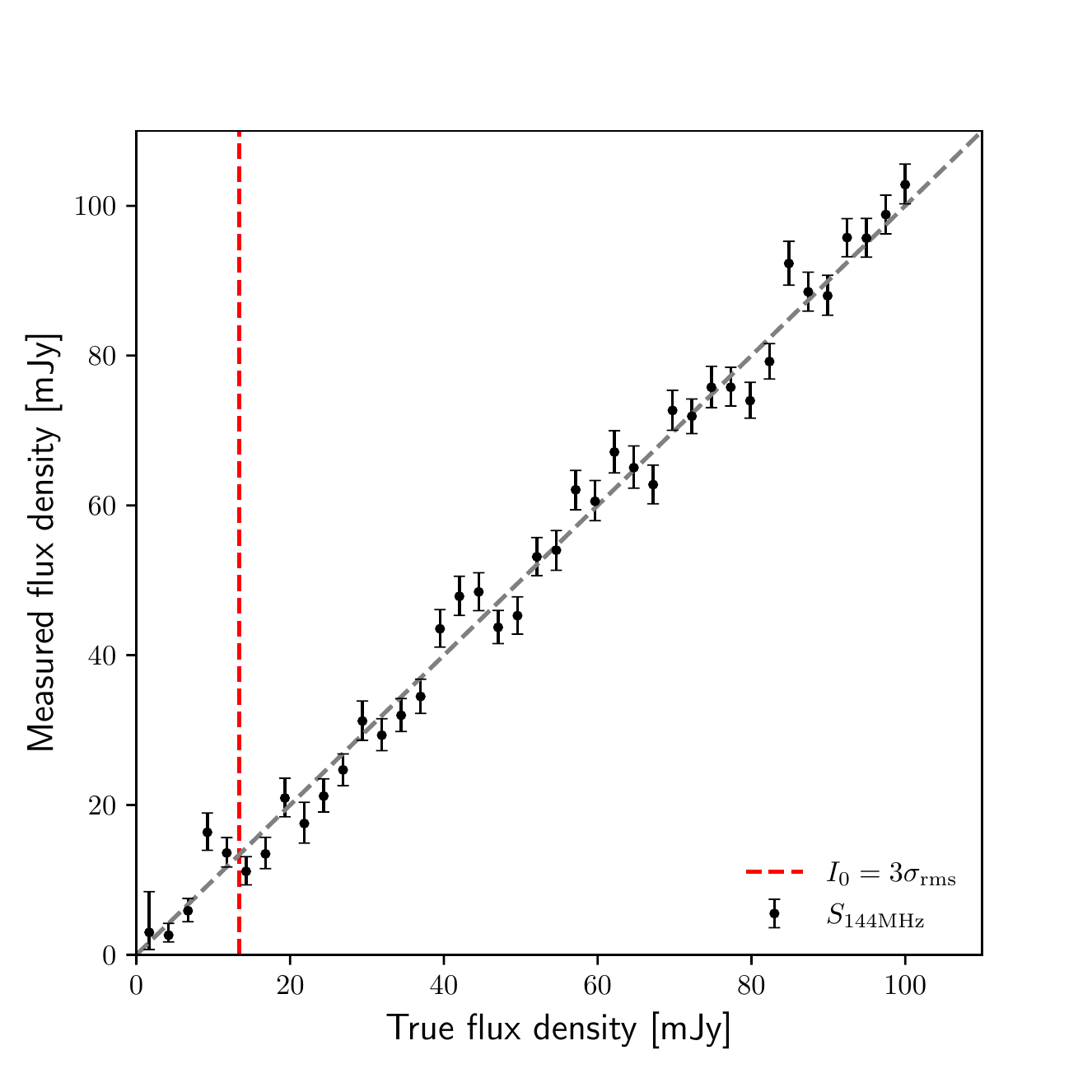}
	\includegraphics[width=0.68\columnwidth,trim={0.1cm 0.2cm 1.cm 0cm},clip]{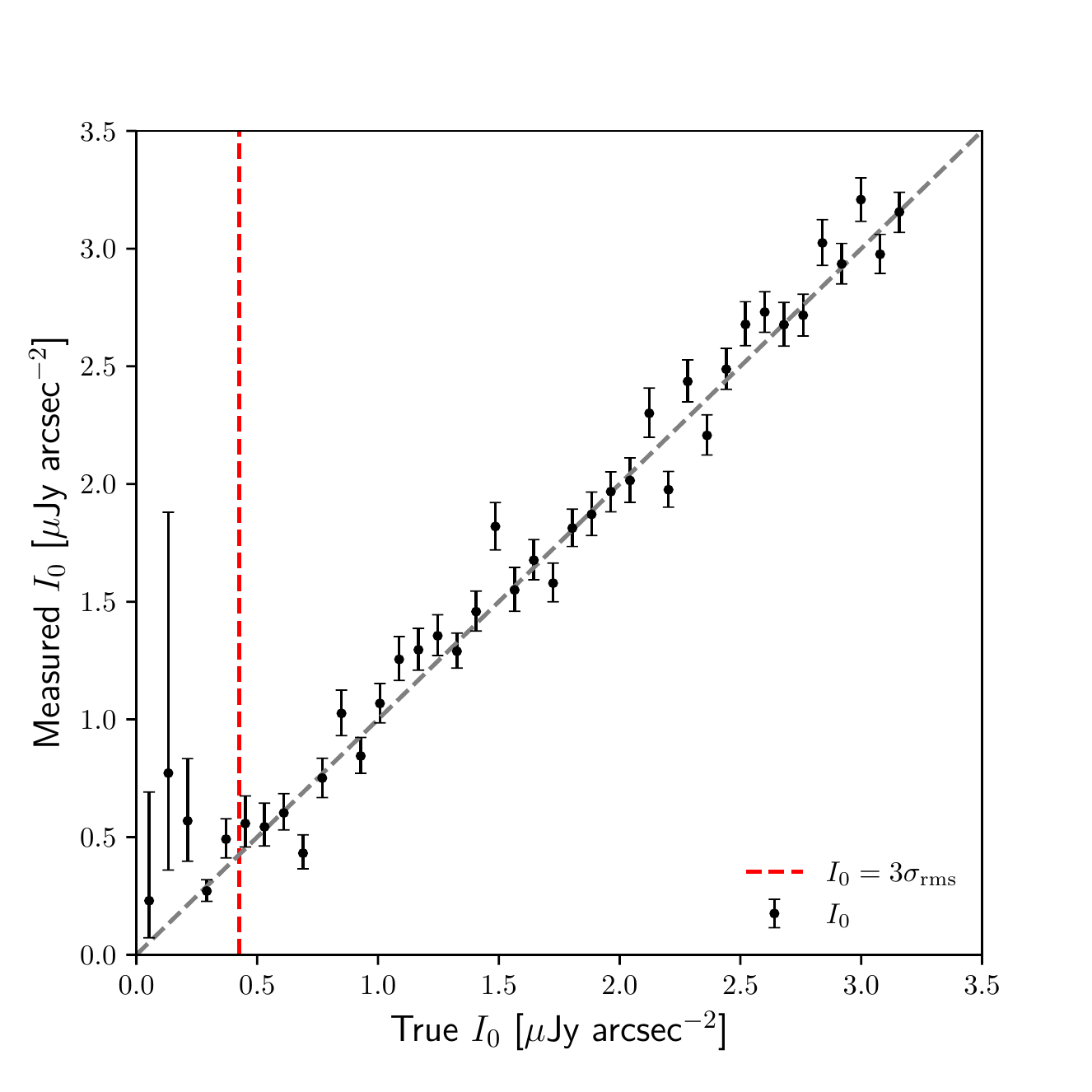}
	\includegraphics[width=0.68\columnwidth,trim={0.cm 0.1cm 1.cm 0cm},clip]{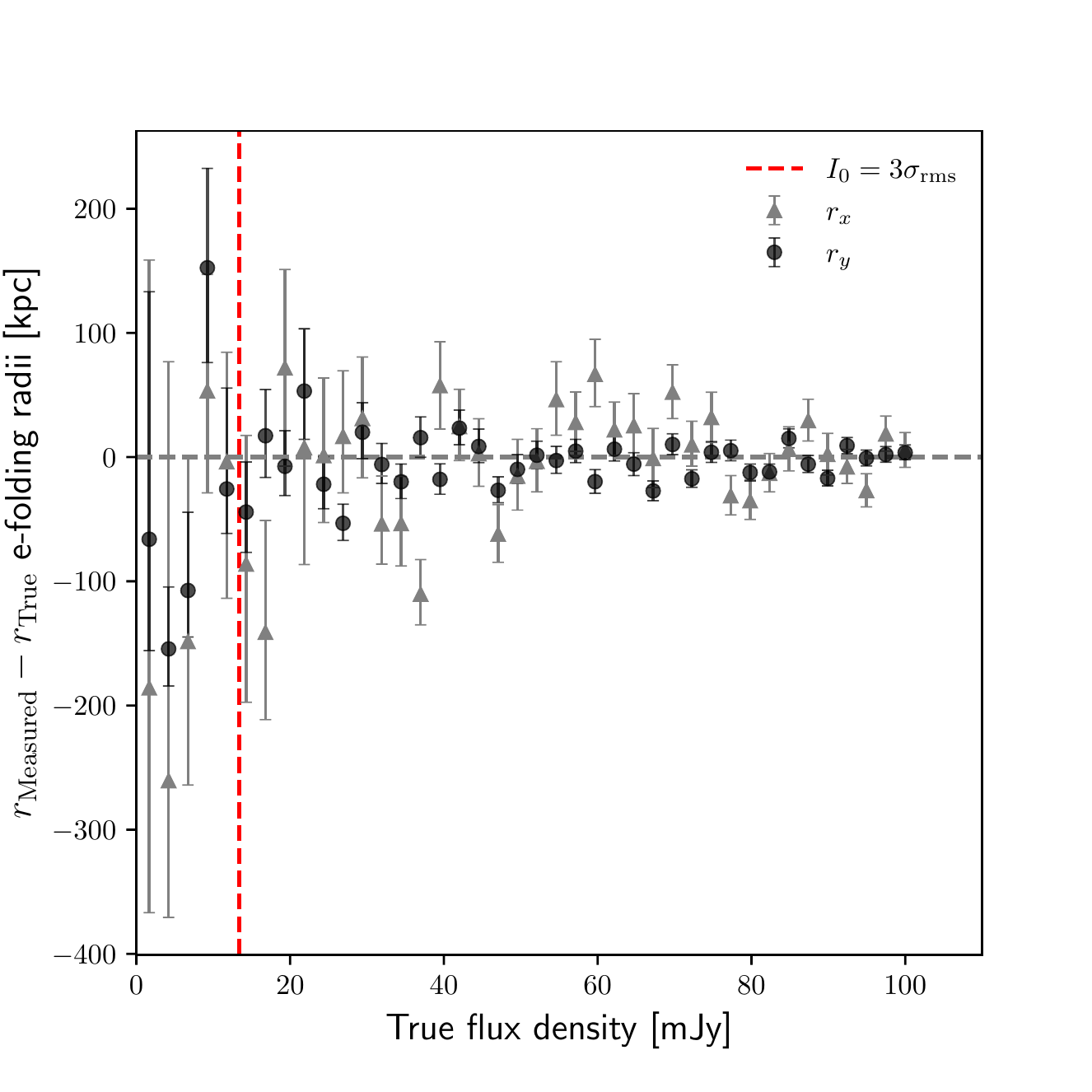}
	
	\caption{\footnotesize \emph{Left panel:} relation between true injected and measured flux density using MCMC. The gray line is the one-to-one optimal relation between the two. Ideally, obtained flux densities need to lie on this line. The red line indicates the point at which the maximum surface brightness $I_0$ drops below $3\sigma_{\text{rms}}$. \emph{Middle panel:} relation between true and measured $I_0$. The accuracy of the obtained $I_0$ decreases significantly below $3\sigma_{\text{rms}}$ and the code starts to overestimate its value. \emph{Right panel:} relation between true injected flux density and obtained e-folding radii minus the true injected radius.}
	\label{fig:artificial-testing}
\end{figure*}

\section{Results, fitting simulated data and examples}
\label{sec:results}

The first part of the results will be dedicated to fitting simulated data, where we will show the performance of the method as a function of surface brightness relative to the noise. This will be investigated by injecting artificially created halos into the algorithm to see how well true parameters can be retrieved. %
In the second part, we will show the application of this algorithm to three radio halos observed with the LOw-Frequency ARray (LOFAR) and the Karl G. Jansky Very Large Array (VLA). We will finally briefly study the influence of different masking on the flux density estimates for a fourth radio halo.

 \begin{figure*}[th]
	\includegraphics[width=0.9\columnwidth,trim={0.1cm 0.2cm 1.cm 0cm},clip]{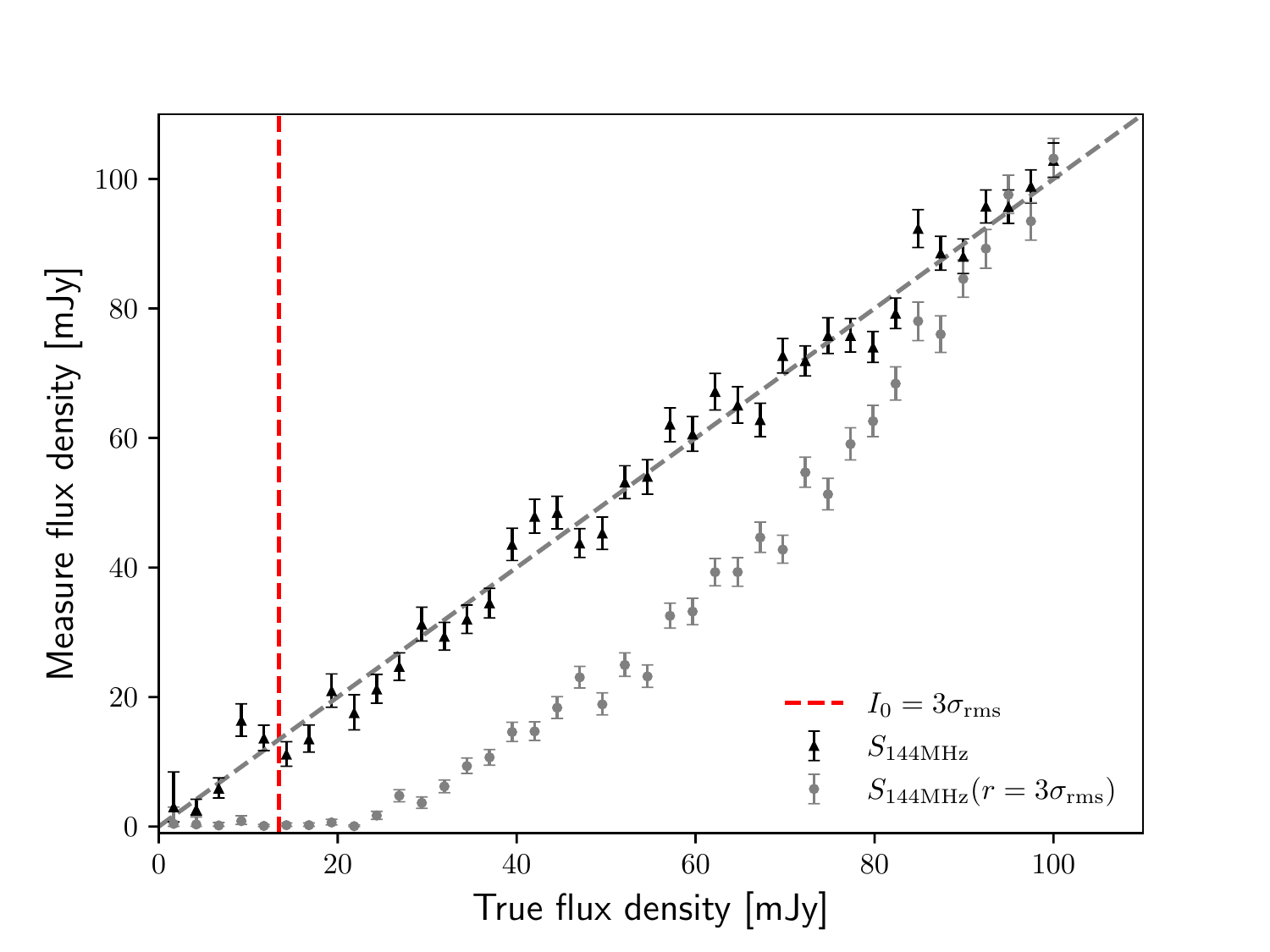}
	\includegraphics[width=0.9\columnwidth,trim={0.1cm 0.2cm 1.cm 0cm},clip]{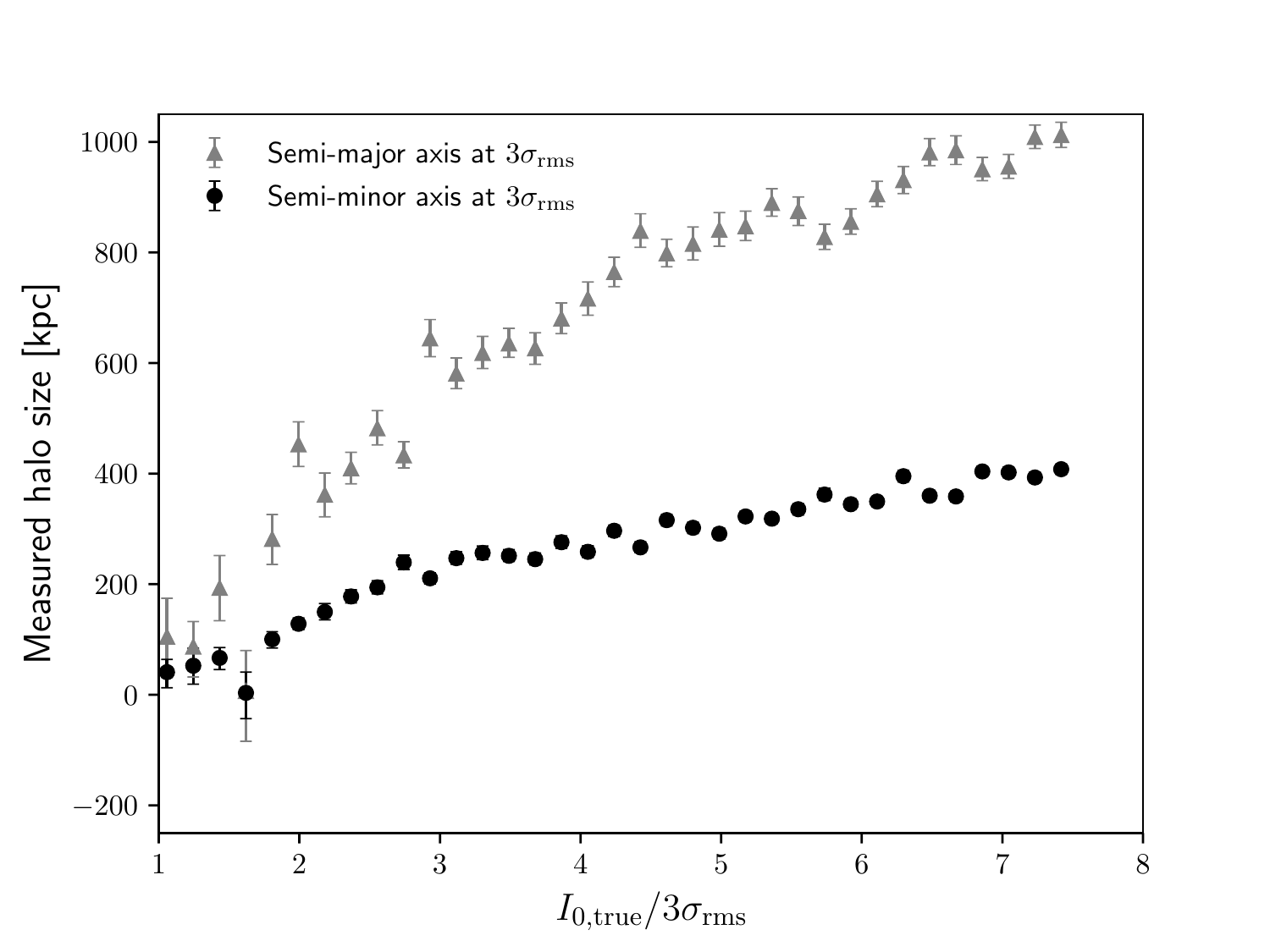}
	
	\caption{\footnotesize  \emph{Left panel:} relation between true injected and measured flux density using MCMC. In this panel, the flux densities with radii corresponding to the $I(\vec{r})=3\sigma_{\text{rms}}$ contour are also shown with grey symbols. This is the method that has often been adopted in the literature. The dashed gray line is the one-to-one optimal relation between the two. The red line indicates the point at which the maximum surface brightness $I_0$ drops below $3\sigma_{\text{rms}}$. So from here $S_{\nu}=0$ for the  $I(\vec{r})=3\sigma_{\text{rms}}$ contour measurements.
	\emph{Right panel:} relation between true injected surface brightness divided by the noise and obtained halo size up to $I(\vec{r})=3\sigma_{\text{rms}}$.}
	\label{fig:artificial-testing-noise}
\end{figure*} 
 
\subsection{Fitting artificially injected halos}
To demonstrate the code is reliable, images with artificially created radio halo sources were used for fitting. The diffuse sources were modelled to be elliptical with a fixed set of parameters. This allows us to check whether the code is able to recover the true parameters of a halo. An artificial halo $A$ is given by $A(\vec{r},\theta)= (I(\vec{r},\theta)+\kappa(\vec{r}))\ast B_{\phi}$ where $\kappa$ is the modelled noise map. The beam parameters are:

$\{b_x,b_y,\phi\} = \{ 32\text{ arcsec} \;, 27\text{ arcsec} \;, 54.2\text{ deg} \}$ \\

and we set the artificial halo at a distance of $z=0.3$.  The map noise is generated each time by making an empty image and filling it with random Gaussian noise $\mathcal{N}(0,1)$. The noise map is then scaled to match $\sigma_{\text{rms}}=0.14$ $\mu$Jy~arcsec$^{-2}$
The total map is then convolved with the known image beam. %
In total, 40 artificial halos of identical shape and varying flux density (or equivalently, varying $I_0$) are constructed to test reproducibility as a function of halo surface brightness. The halo that is injected is based on a rotated elliptical model with parameters: 
\begin{align*}
	&r_x= 500\;\text{kpc},\\
	&r_y= 200\;\text{kpc},\\
	&\phi = 2\pi/3 \mbox{.}
\end{align*}

$I_0$ varies based on the flux density. Flux densities range from $1.67 - 100$ mJy ($I_0$ is then calculated via Equation~\eqref{6D-model}). Modelled halos corresponding to $S_{\nu}=11.75$ mJy and $S_{\nu}=100$ mJy are shown in Figure~\ref{fig:artificial_example}. 

The flux density, $I_0$ and radii values obtained running the algorithm are presented in Figure~\ref{fig:artificial-testing}. It shows that the code is behaving as expected: it is able to recover well the flux densities for the high surface brightness halos, while it fails to settle at the 4 lowest flux density values. This is not surprising since the maximum surface brightness of these halos is well below the $3\sigma_{\text{rms}}$ level. Interestingly, the figure shows accurate flux density estimates at low surface brightness despite the high uncertainties of the other parameters. This is possibly due to a correlation between maximum surface brightness and radii. This is also seen in Figure~\ref{fig:artificial-testing}, where at low surface brightness, radii seem to be underestimated (leading to lower flux densities) while $I_0$ is being overestimated (leading to higher flux densities) resulting in an accurate flux density where both effects are canceled out.

Conventionally, sizes of radio halos are measured visually or up to $3\sigma_{\text{rms}}$ contours \citep[e.g.][]{2007MNRAS.378.1565C}. These noise-based size estimations dramatically influence the measured flux density. Figure~\ref{fig:artificial-testing-noise} stresses the inaccuracies resulting from this. The left panel shows both the flux estimated as in Figure~\ref{fig:artificial-testing} in black and the flux density with a radius corresponding to $3\sigma_{\text{rms}}$. It is clear that this estimation is accurate for very bright halos but underestimates the flux for faint halos and it naturally drops to zero flux as the maximum surface brightness reaches $3\sigma_{\text{rms}}$. The right panel shows the halo size (up to the $3\sigma_{\text{rms}}$ contour) for all the artificial halos. The figures illustrate how basing results on manual/visual measurements can bias the outcome. 

 \begin{figure*}[t]
	\includegraphics[width=0.66\columnwidth,trim={0.1cm 1.2cm 3.5cm 0cm},clip]{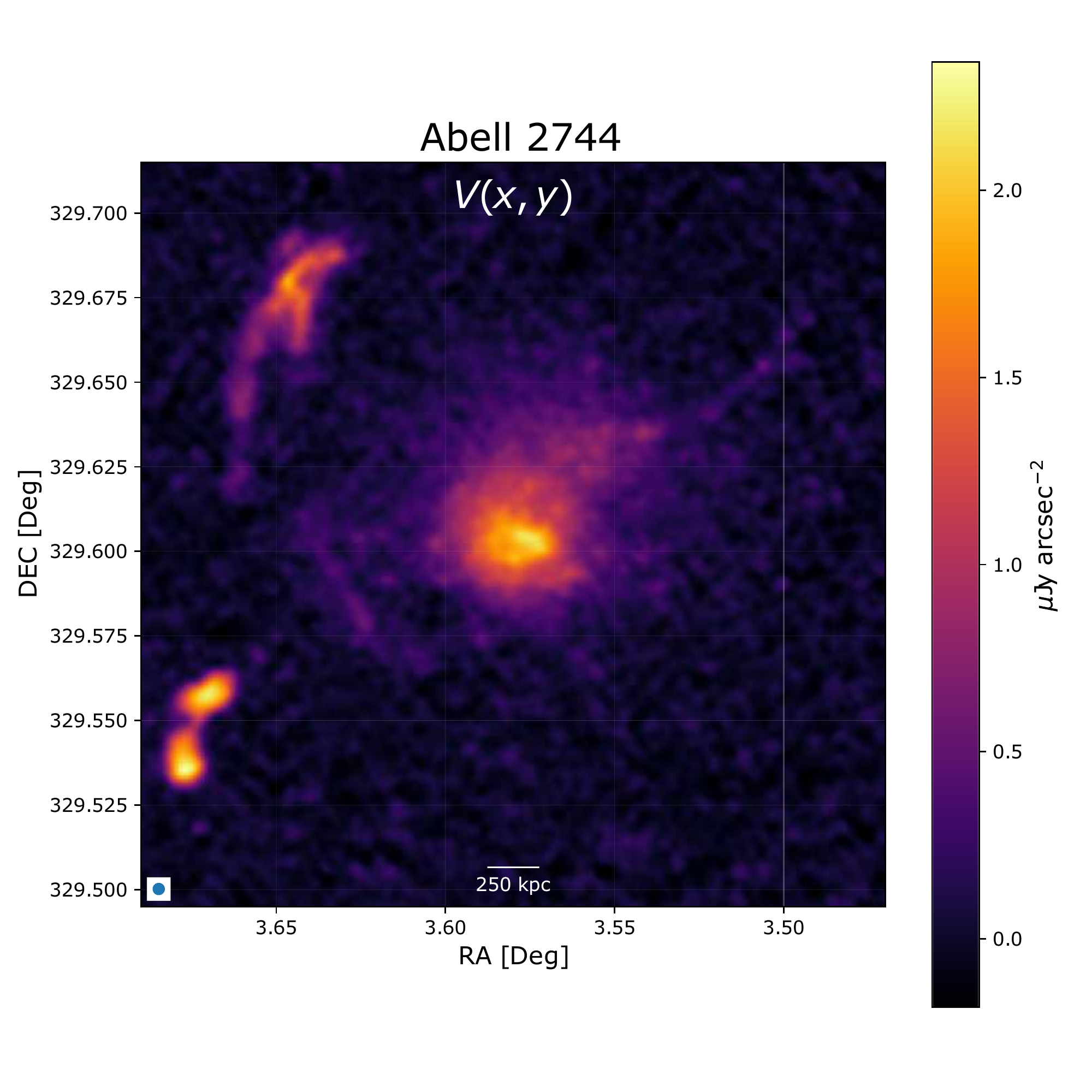}
	\includegraphics[width=0.66\columnwidth,trim={0.1cm 1.2cm 3.5cm 0cm},clip]{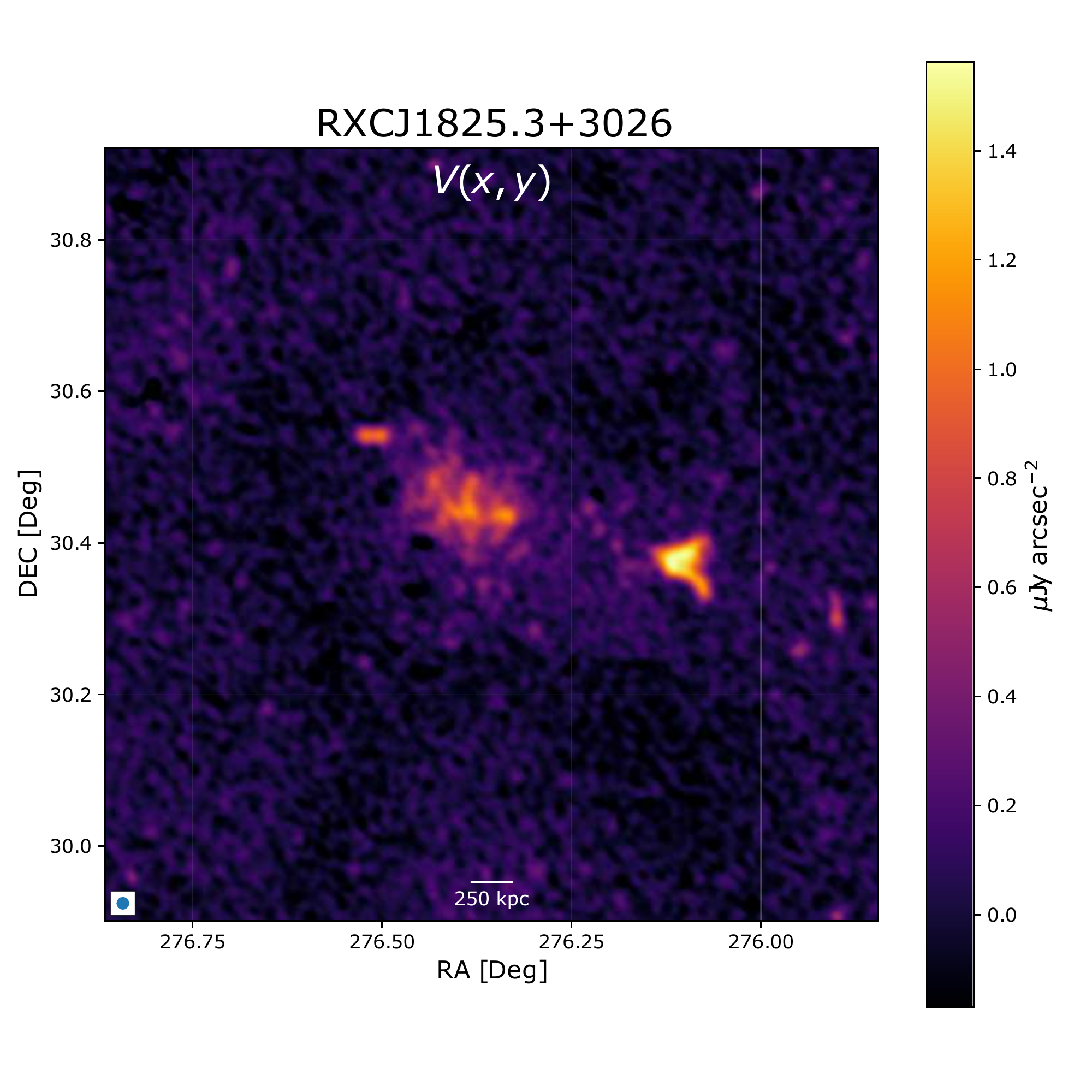}
	\includegraphics[width=0.66\columnwidth,trim={0.1cm 1.1cm 3.5cm 0cm},clip]{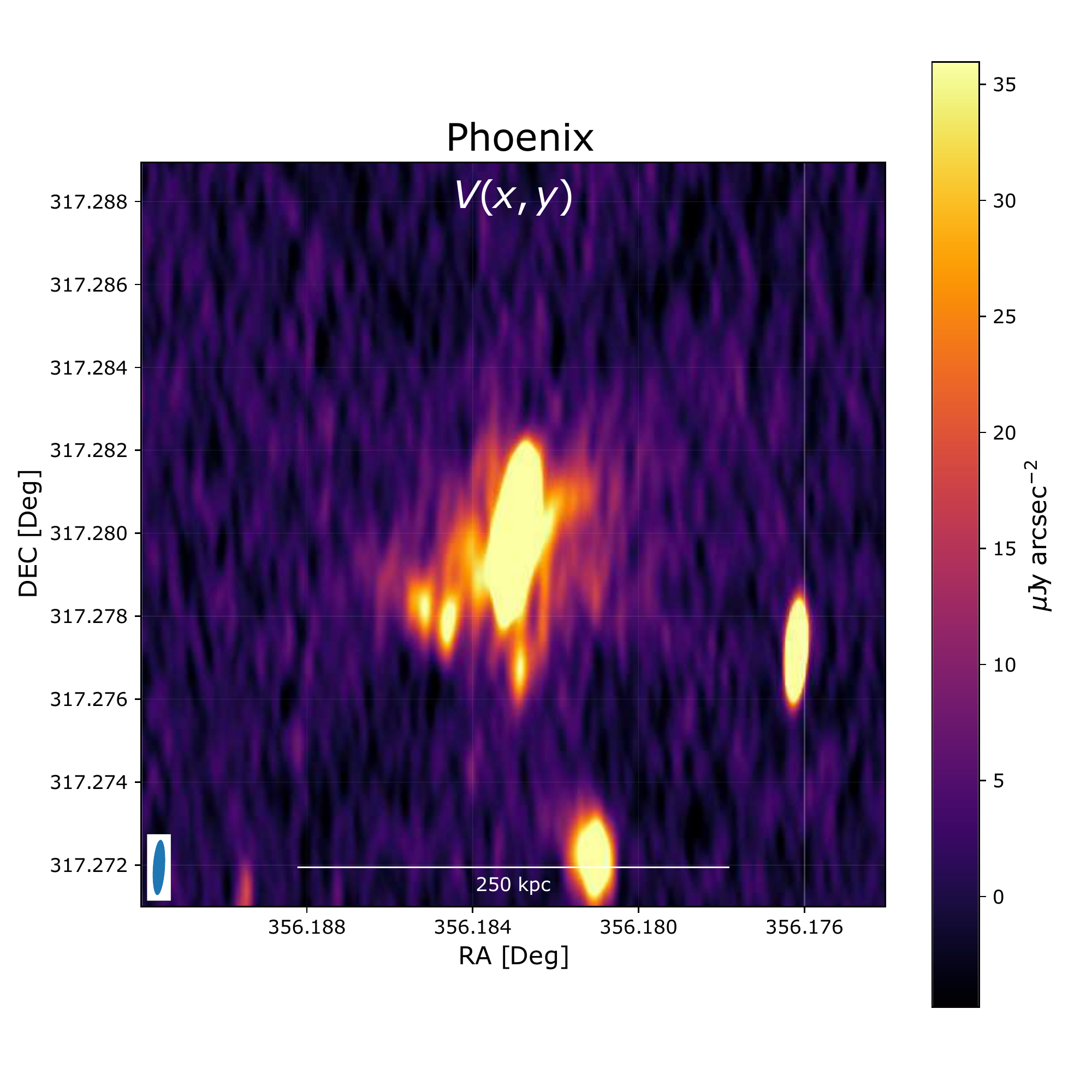}
	
		\caption{\footnotesize Images of the three clusters used in this work. From left to right: Abell\,2744, RXC\,J1825.3+3026 and the Phoenix cluster. Colour bars are left out here but can be found in Figures \ref{fig:Abell_all}, \ref{fig:RXC_all} and \ref{fig:phoenix_all} respectively.}
	\label{fig:plain_data}
\end{figure*}

 \begin{figure*}[t] %
	\includegraphics[width=\linewidth]{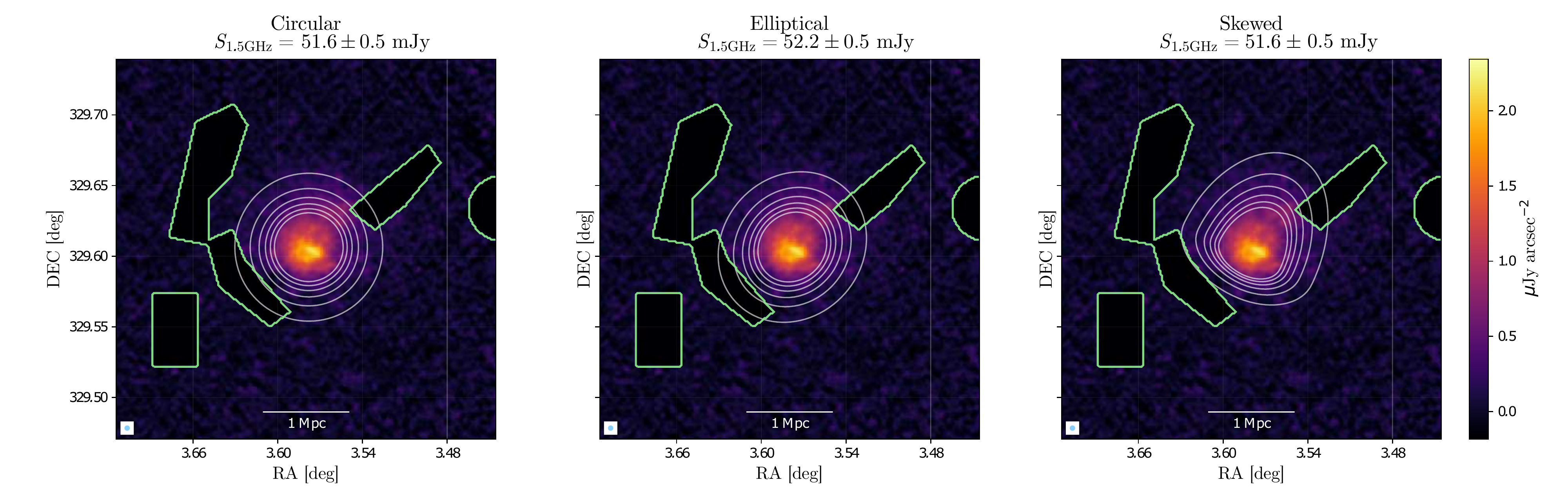}
		\caption{\footnotesize Fitting result for compact source subtracted image of \emph{Abell\,2744} at 1.5~GHz. Physical scale and image beam are depicted on the bottom. The contours show the model at $[1,2,...,7]\times\sigma_{rms}$ levels. Contaminated regions are masked out and contoured in green. Estimated fluxes are indicated on the top. All relevant results are summarised in Tab. \ref{tab:results}.}
	\label{fig:Abell_all}
\end{figure*}

\subsection{Three examples}
\label{sec:examples}
We now use two giant radio halos (Abell\,2744 and  RXC\\*J1825.3+3026) and a radio mini-halo (Phoenix Cluster) with known flux densities as examples to show how {\tt Halo-FDCA} compares to manual measurement methods.

The radio halo in Abell\,2744 was first identified by \citet{1999NewA....4..141G} and the flux density has been estimated by \citet{2017ApJ...845...81P} using 1.5~GHz VLA images. \\* RXC\,J1825.3+3026 is a cluster at $z=0.065$ with a $M_{500}$ mass of $4.08 \pm 0.13$ M$_{\odot}$ belonging to the Lyra complex \citep{2019A&A...632A..27C, 2020A&A...633A.108G}. The flux density has been estimated by \citet{2019A&A...630A..77B} using 144~MHz LOFAR observations. The Phoenix cluster is a high redshift ($z=0.596$) massive ($M_{500}= 12.6^{+2.0}_{-1.5}\times10^{14}$ M$_{\odot}$) relaxed galaxy cluster with a mini-halo at its centre \citep{2012Natur.488..349M,2014ApJ...786L..17V}. Recently, \citet{2020arXiv200913238T} studied this system using deep multi-frequency multi-configuration VLA observations. In the following, we will be using the same images as published in the mentioned works. We note that the uncertainty presented here only takes into account uncertainties introduced by the fitting process and we ignore flux-scale uncertainties.

Figure \ref{fig:plain_data} shows all three clusters as they were used during fitting. These figures show the contamination due to compact sources or partially subtracted extended radio galaxies that are left out before a mask is applied during fitting. An important aspect of the fitting is checking whether there is a statistically significant result. The figures and table shown in this section help motivate {\tt Halo-FDCA} converged on the expected emission. %
The signal-to-noise ratio, $S/N$ (defined as the flux density value divided by the uncertainty) and $\chi^2_{\text{red}}$ provide a handle on what model is best to use. The skewed model might describe bright, extended, or irregular halos better (especially with deep, high signal halo images) while the circular model is a safe choice for faint halos. %

We performed the fitting on a local 96 core (four AMD EPYC 7401 24-Core, 2.00~GHZ CPUs) node with a total of 512~GB of internal memory. %
For fitting Abell\,2744 (384$\times$384 pixels$^2$) for instance, the code took 20 minutes to run. This is with multiprocessing turned on. Running the code without using a Python parallelization increases run-times to over 2.5 hours. %

\subsubsection{Abell\,2744}
For this object, a 1.5~GHz VLA image was used. The initial image was 384$\times$384 pixels$^2$ with scale $4$ arcsec~pixel$^{-1}$. Beam characteristics for this observation are $b_x=13,23$ arcsec, $b_y=13,23$ arcsec and $\phi=-80.77$ deg. The noise of the original image is $\sigma_{\text{rms}}=18.10$ $\mu$Jy beam$^{-1}$. Due to regridding, the uncorrelated image had a size of 60$\times$60 pixels$^2$ with $\sigma_{\text{rms}}=15.44$ $\mu$Jy beam$^{-1}$. We fitted three models to the image: the circular, rotated elliptical and skewed model. The general results for all three models are shown in Figure~\ref{fig:Abell_all}. Accompanying this figure, Table~\ref{tab:results} provides numerical results.

\begin{figure*}[hbt!]
	\includegraphics[width=\linewidth,trim={0.5cm 0.5cm 0.5cm 0cm},clip]{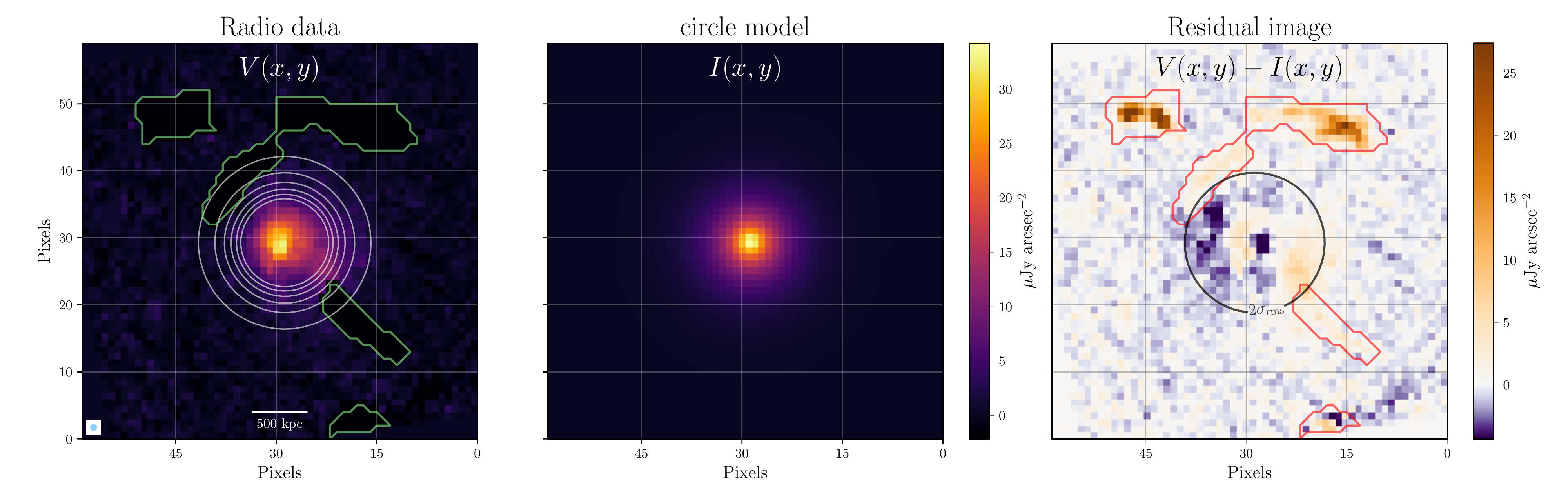}
	\includegraphics[width=\linewidth,trim={0.5cm 0.5cm 0.5cm 0cm},clip]{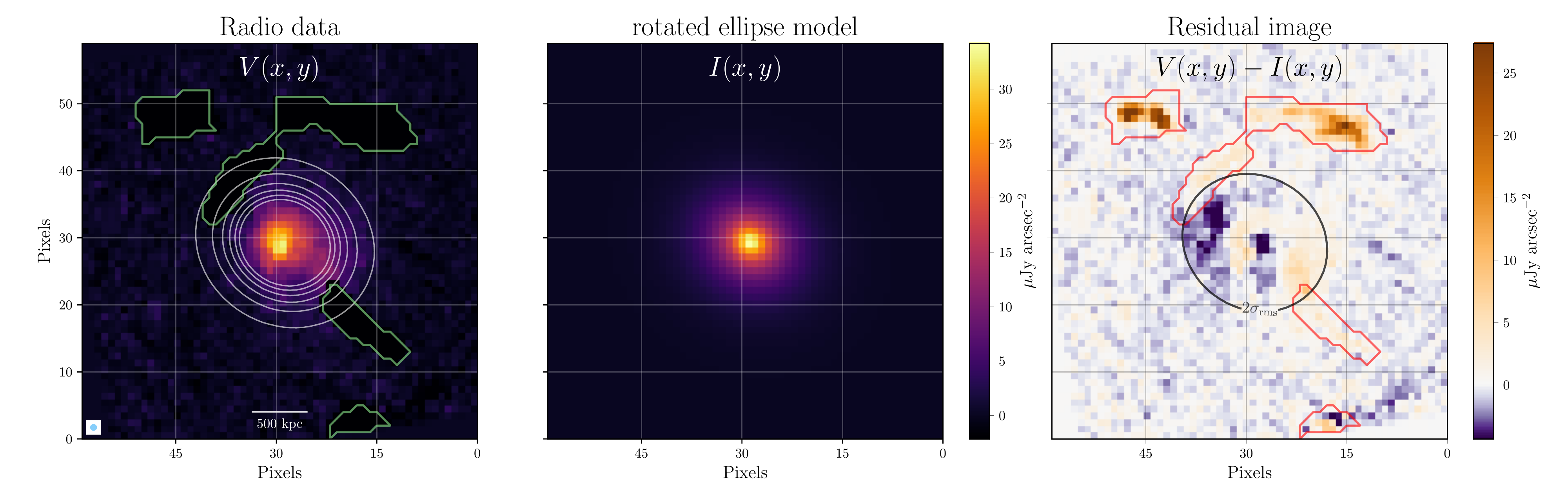}
	\includegraphics[width=\linewidth,trim={0.5cm 0.5cm 0.5cm 0cm},clip]{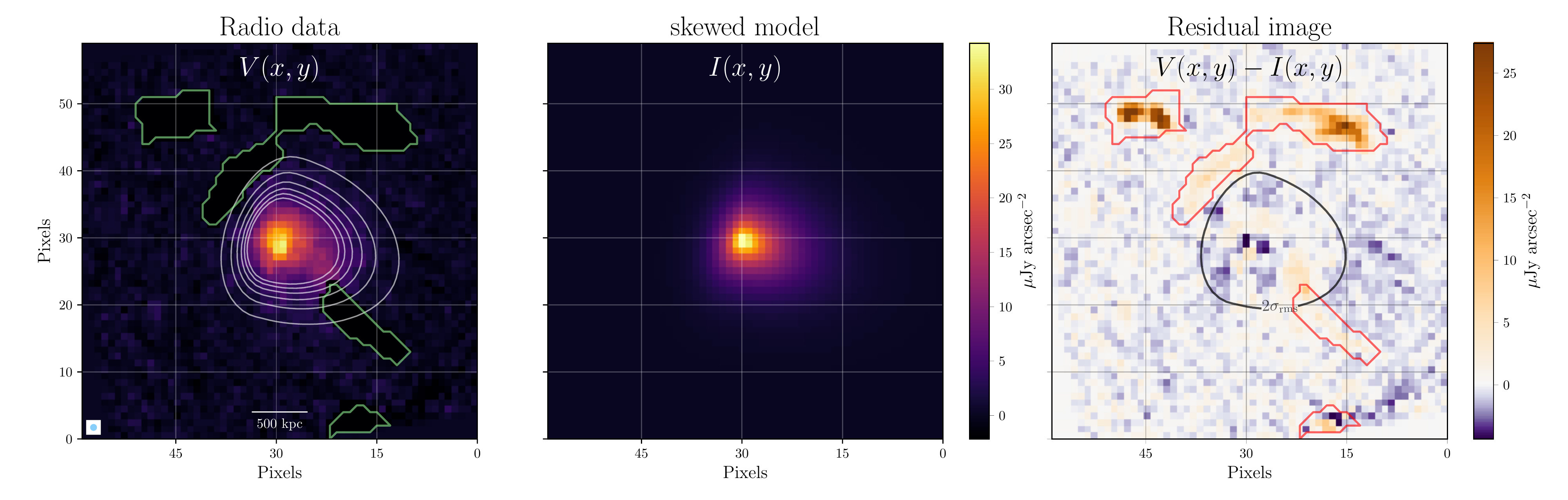}
		\caption{\footnotesize Fitting result for \emph{Abell\,2744} at 1.5~GHz. Models from top to bottom are circle, rotated ellipse \& skewed. \emph{Left panel:} Regridded image for all three models. The physical scale and image beam are depicted below. The contours show the model at $[1,2,...,7]\times\sigma_{rms}$ levels. Contaminated regions are masked out and contoured in green. \emph{Middle panel:} Model map \emph{Right panel:} residual image. %
		The contour shows the $2\sigma_{rms}$ level of the model. The red contour now shows the masked regions, but this time the contamination is visible.}
	\label{fig:abell_models}
\end{figure*}

Table~\ref{tab:results} shows that all three models result in almost identical flux density values within one standard deviation, which suggest no preference for any of the three models regarding the flux density. The same is true for the maximum surface brightness $I_0$ values. Based on the $\chi^2_{\text{red}}$ value, the skewed model is favored over the other two, the skewed model is also the one with the smallest relative uncertainty. 

\begin{figure*}[hbt!]%
	\includegraphics[width=\linewidth]{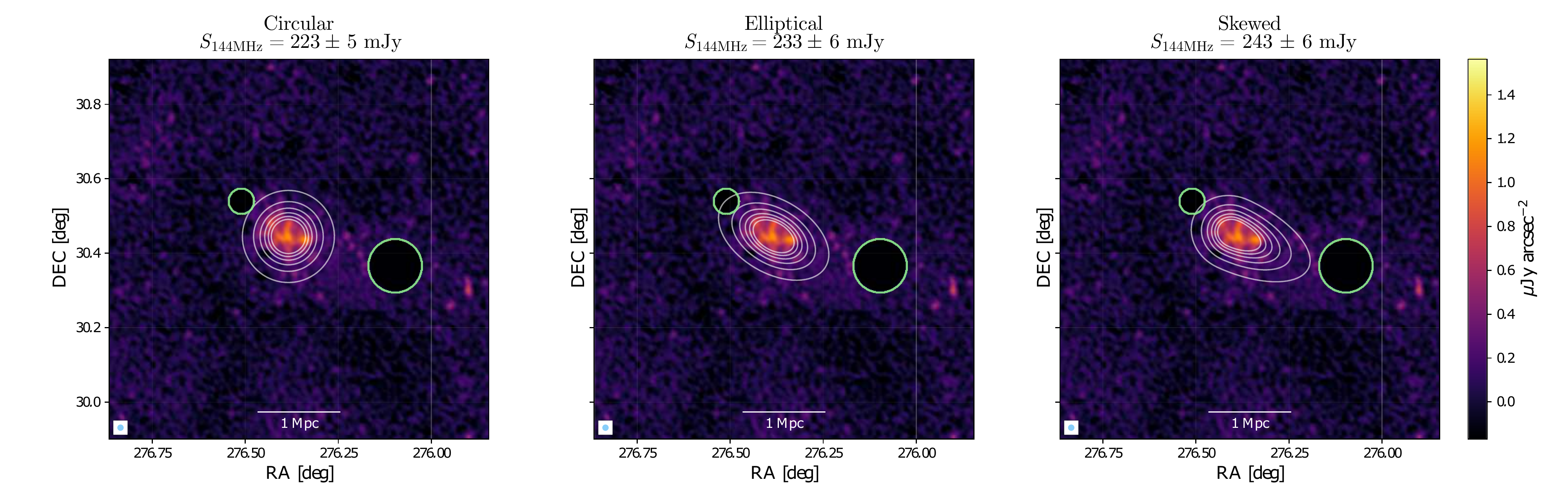}
		\caption{\footnotesize Fitting result for point source subtracted image of \emph{RXC\,J1825.3+3026} at 144~MHz. Physical scale and image beam are depicted in the bottom corner. The contours show the model at $[1,2,...,7]\times\sigma_{rms}$ levels. Contaminated regions are masked out and contoured in green. The estimated fluxes densities are indicated on top of the figures. All relevant results are summarised in Tab. \ref{tab:results}.}
	\label{fig:RXC_all}
\end{figure*}

\begin{table*}[b]
\begin{center}
\begin{tabular}{lccccccc}
\toprule
Cluster & $S_{\nu}$  & $I_0$ & $r_{\text{max}}$  %
&$\chi^2_{\text{red}}$& $S_{\nu}/\sigma_{S_{\nu}}$ &Figure\\
& mJy  & $\mu$Jy~arcsec$^{-2}$ & kpc & &%
& \\
		\midrule
Abell\,2744 & $51.61^{+0.49}_{-0.48}$ & $2.60^{+0.02}_{-0.02}$  & $255^{+2}_{-2}$& %
$1.486$ & $106.52$ & \ref{fig:abell_models} \vspace{0.05cm}\\
 & $52.17^{+0.50}_{-0.49}$ & $2.58^{+0.02}_{-0.02}$  & $273^{+3}_{-3}$& %
 $1.461$& $104.56$& \vspace{0.05cm}\\
 & $51.62^{+0.47}_{-0.46}$ & $2.68^{+0.02}_{-0.02}$  & $372.0^{+5}_{-5}$& %
 $1.113$& $111.50$& \vspace{0.1cm}\\

RXC\,J1825.3+3026 & $223.00^{+5.46}_{-5.46}$ & $1.47^{+0.04}_{-0.04}$  & $194^{+4}_{-4}$& %
$1.199$ & $40.86$ & \ref{fig:rxc_models} \vspace{0.05cm}\\
 & $233.09^{+5.91}_{-5.71}$ & $1.49^{+0.04}_{-0.04}$  & $254^{+7}_{-7}$& %
 $1.126$& $40.13$& \vspace{0.05cm}\\
 & $243.40^{+6.39}_{-6.30}$ & $1.43^{+0.04}_{-0.04}$  & $360.2^{+20}_{-19}$& %
 $1.108$& $38.36$& \vspace{0.1cm}\\

Phoenix & $8.15^{+0.18}_{-0.18}$ & $1.32^{+0.13}_{-0.13}$ & $27.6^{+0.7}_{-0.7}$ & 1.05 & 46.06 & \ref{fig:phoenix_models} \vspace{0.05
cm}\\
 & $8.39^{+0.17}_{-0.16}$ & $1.36^{+0.14}_{-0.14}$ & $30.7^{+0.9}_{-0.8}$ & 0.94  & 48.63& \vspace{0.05cm}\\
 & $8.15^{+0.18}_{-0.18}$ & $1.32^{+0.13}_{-0.13}$ & $39^{+2}_{-2}$ &  0.97 & 46.35& \vspace{0.1cm}\\
\bottomrule
\end{tabular}
\end{center}
	\caption{\footnotesize Most notable results for fitting different models to halo images. Uncertainties on presented values are taken from MCMC walker distributions. Per cluster, results are presented for the circular, elliptical and skewed model respectively. \emph{Col 2}: Halo flux densities at $\nu=1.5$~GHz for Abell\,2744 and the Phoenix cluster and $\nu=144$~MHz for RXC\,J1825.3+3026, \emph{Col 3}: Amplitude parameter as mentioned in all models, \emph{Col 4}: maximum $e$-folding radius per model to indicate the halos estimated size (the skewed model for instance has four $r$ parameters), \emph{Col 5}: reduced $\chi$ squared value, \emph{Col 6}: Signal to noise based on flux value and its error, \emph{Col 7}: Cluster image with fit overlay (Appendix \ref{app:results}).}
\label{tab:results}
\end{table*} 

More detailed figures for every model are shown in Fig. \ref{fig:abell_models} also points to the skewed model as the best description of this halo since the residual image for the bottom row shows the least amount of over-subtraction by the model within the $2\sigma_{\text{rms}}$ region. This under- or over-subtraction is mainly caused by cluster substructure or possible inaccurate subtraction of discrete sources embedded in the diffuse emission. The residual images are thus potentially useful for studying substructure. 

The residual images suggest that the last model provides a more accurate description of the physical object and we thus conclude that the flux density of this radio halo is \\*$S_{\text{1.5 GHz}} = 51.6\pm0.5$ mJy with $e$-folding radii
\begin{align*}
    r_e&=\{r_{x^+},r_{x^-},r_{y^+},r_{y^-}\}\\
    &=\{372\pm5,\;150\pm 5,\;256\pm6,\;228\pm7\}\;\text{kpc}.
\end{align*} 
Visually, the halo extends up to approximately two times the obtained radii.

The flux density is slightly above measurements from \citet{2017ApJ...845...81P}, who found a flux density of $S_{\text{1.5 GHz}} = 45.1\pm2.3$ mJy. They manually drew a circular region around the emission and integrated the emission within that. The higher value reported here is probably due to the model being extrapolated to infinity well below the noise level and not being cut based on visual appearance. This causes the value to be about 20\% higher than when using $3r_e$ as the integration limit. See the Discussion for some more comments on the choice of integration radii. Furthermore, in \citet{2017ApJ...845...81P} the flux density in the masked regions was not extrapolated using, contrary to what is done by \texttt{Halo-FDCA}.

\begin{figure*}[hbt!]%
	\includegraphics[width=\linewidth,trim={0.5cm 0.5cm 0.5cm 0cm},clip]{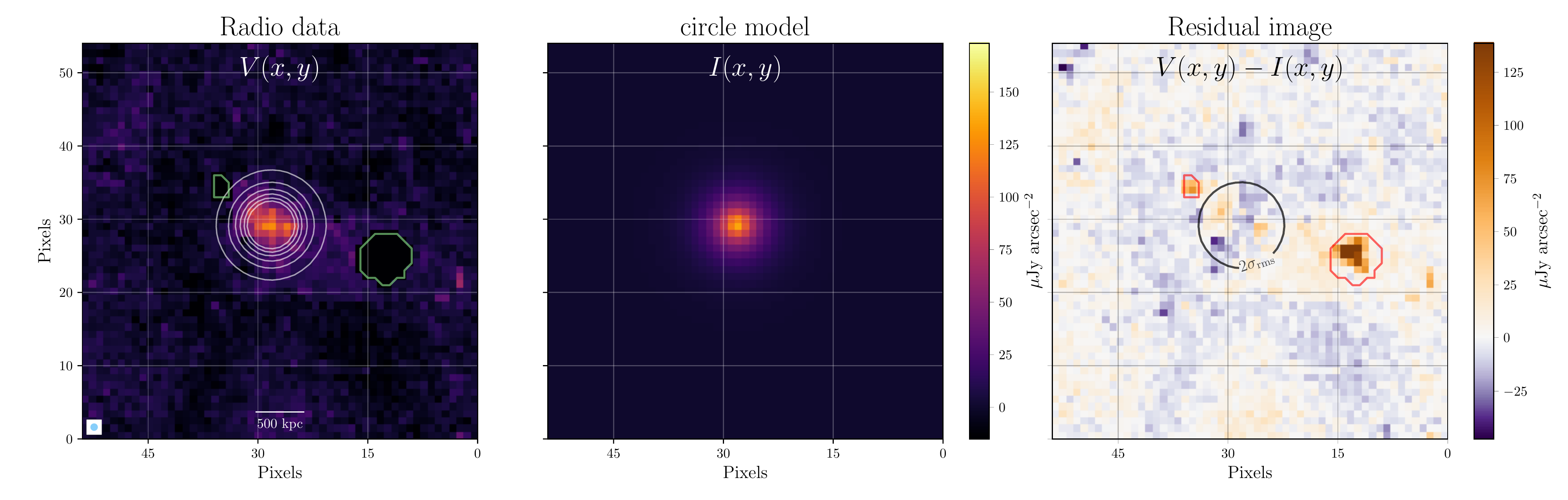}
	\includegraphics[width=\linewidth,trim={0.5cm 0.5cm 0.5cm 0cm},clip]{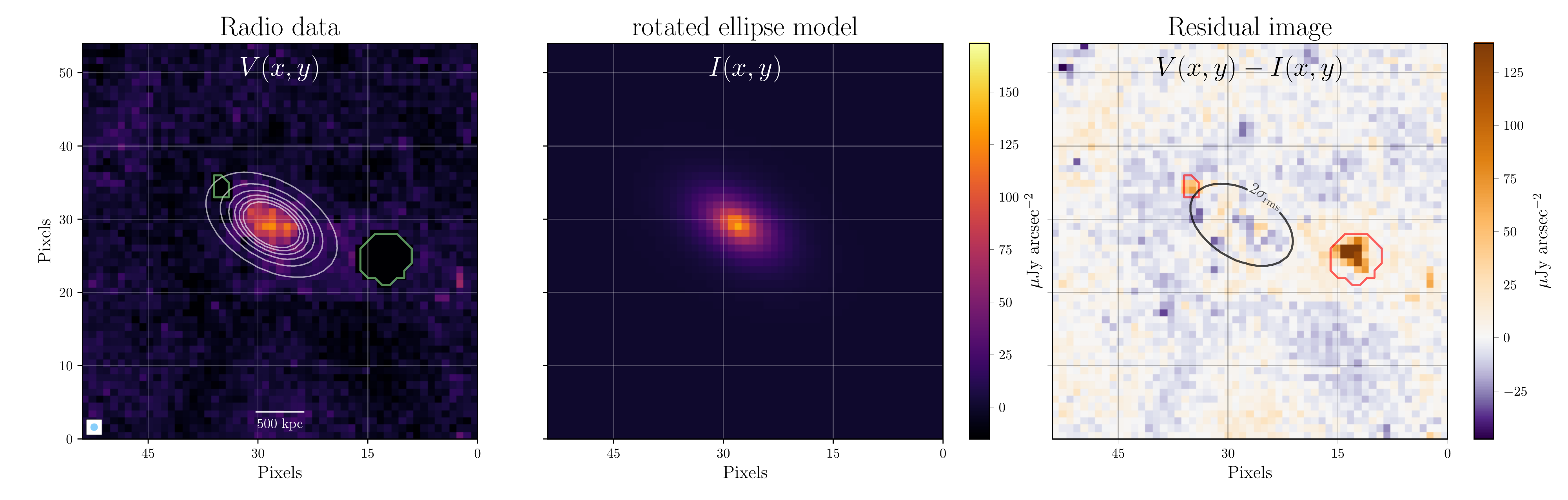}
	\includegraphics[width=\linewidth,trim={0.5cm 0.5cm 0.5cm 0cm},clip]{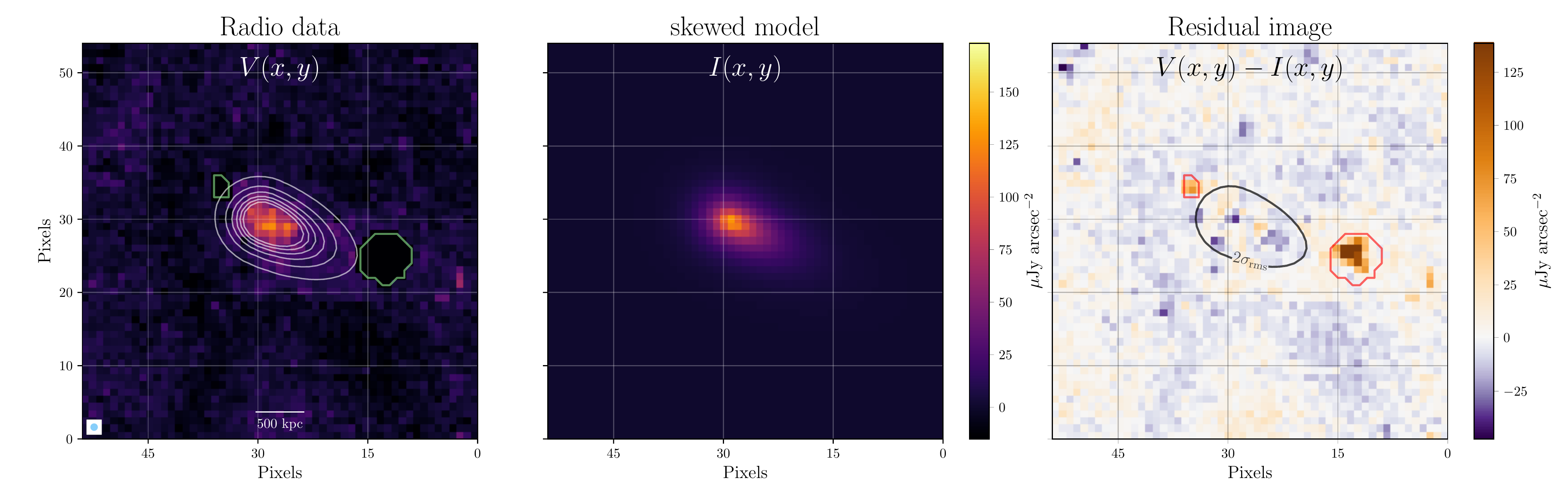}
		\caption{\footnotesize Same as Fig. \ref{fig:abell_models}. Fitting result for \emph{RXC\,J1825.3+3026} at 144~MHz.}
	\label{fig:rxc_models}
\end{figure*}

\subsubsection{RXC\,J1825.3+3026}
For this object, LOFAR 144~MHz observations, published in \citet{2019A&A...630A..77B} are used. The initial image is 613$\times$613 pixels$^2$ large with a  $6$~arcsec~pixel$^{-1}$ scale. The regridded image size is 55$\times$55 pixels$^2$ with $\sigma_{\text{rms}}=277$ $\mu$Jy beam$^{-1}$. This regridded image was obtained using $b_x=b_y=1$ arcmin and $\phi=0$ as beam characteristics. The model reconstruction of the radio halo for the circle, ellipse and skewed model are shown in Figure~\ref{fig:RXC_all}. The numerical results for these fits are again shown in Table~\ref{tab:results}.

Figure~\ref{fig:RXC_all} shows the algorithm suggesting quite elongated radio emission in RXC\,J1825.3+3026. The skewed model even points to the existence of a low surface brightness extension in the lower right corner of the radio halo. Looking at the $\chi^2_{\text{red}}$ value in Table~\ref{tab:results}, this situation is preferable, even though it has the highest relative uncertainty. From the table, it is clear that the estimated flux density varies significantly per chosen model. %

The radio halo RXC\,J1825.3+3026 is known to show a low surface brightness extension towards the SE direction \citep{2019A&A...630A..77B}. Therefore, the elongated shape found by the skewed model is not surprising. We found a flux density: $S_{\text{144 MHz}}=243\pm6$ mJy with radii $r_e=\{180\pm17,\;360\pm20,\;159\pm10,\;155\pm10\}$ kpc. A more conservative estimate, by focusing on the circle model only would be $S_{\text{144 MHz}}=223\pm6$ mJy with $r_e=194\pm4$ kpc. 

This value is higher than reported by \citet{2019A&A...630A..77B} ($S_{144}=163\pm47$ mJy), who however reported the flux density of the radio halo from a rectangular region roughly defined by the 3 sigma level contours in the cluster center, excluding its SE extension. Our code demonstrates it can recover low-surface brightness emission that may contribute to the final flux density of the halo. 

\begin{figure*}[hbt!]%
	\includegraphics[width=\linewidth]{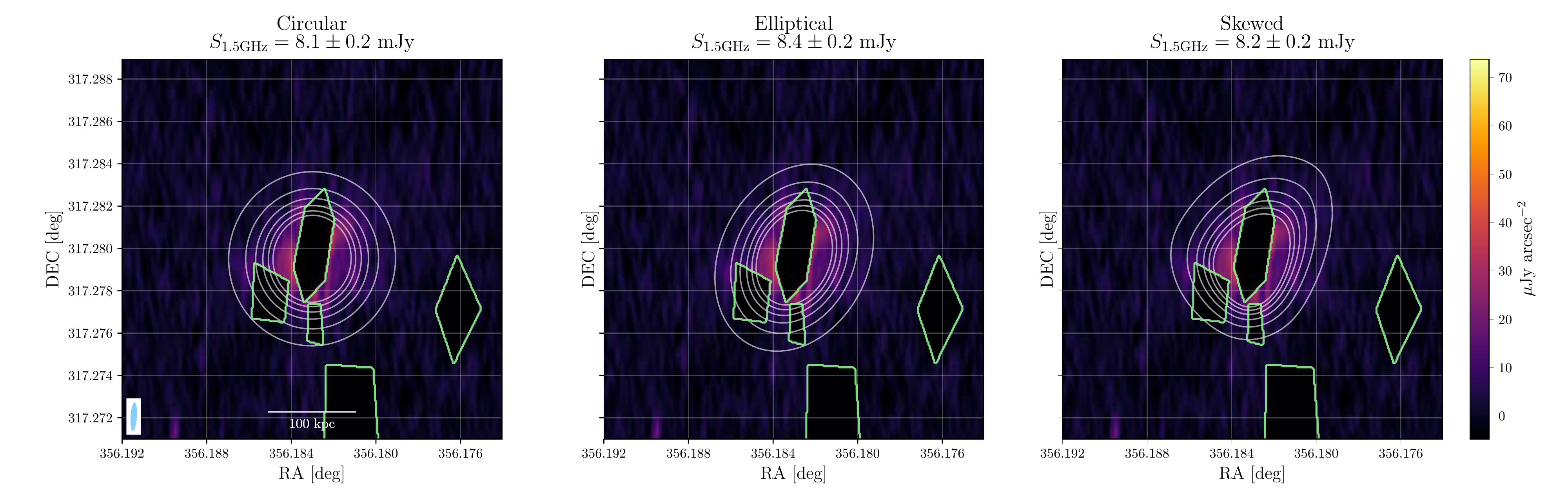}
		\caption{\footnotesize Fitting result for the VLA image of the Phoenix cluster at 1.5~GHz. The image beam is depicted on the bottom. It can be seen that the beam is very elongated. The contours show the model at $[1,2,...,7]\times\sigma_{rms}$ levels. Contaminated regions are masked out and contoured in green. The estimated fluxes densities are indicated on top of the figures. All relevant results are summarised in Tab. \ref{tab:results}.}
	\label{fig:phoenix_all}
\end{figure*}

\subsubsection{Phoenix cluster}
\label{sec:phoenix}
For the last example, we use an image of the mini-halo located in the Phoenix cluster, first reported by \citet{2014ApJ...786L..17V} using 610~MHz GMRT data. Here we use 1.5~GHz VLA data from \citet{2020arXiv200913238T}. The initial image is 324$\times$324 pixels$^2$ with a noise of $\sigma_{\text{rms}}=13.57$ $\mu$Jy beam$^{-1}$. The image has a $0.2$ arcsec~pixel$^{-1}$ to sky scale. During the fitting, an image of 11$\times$52 pixels$^2$ with $10.42$ $\mu$Jy beam$^{-1}$ was used. This image is rather rectangular due to the elongated image beam with $b_x=4,82$ arcsec by $b_y=1,04$ arcsec and $\phi=176.4$ deg. The results for all models are shown in Figure~\ref{fig:phoenix_all} and the fitted parameters are again listed in Table~\ref{tab:results}.

\begin{figure*}[hbt!]%
	\includegraphics[width=1.05\linewidth]{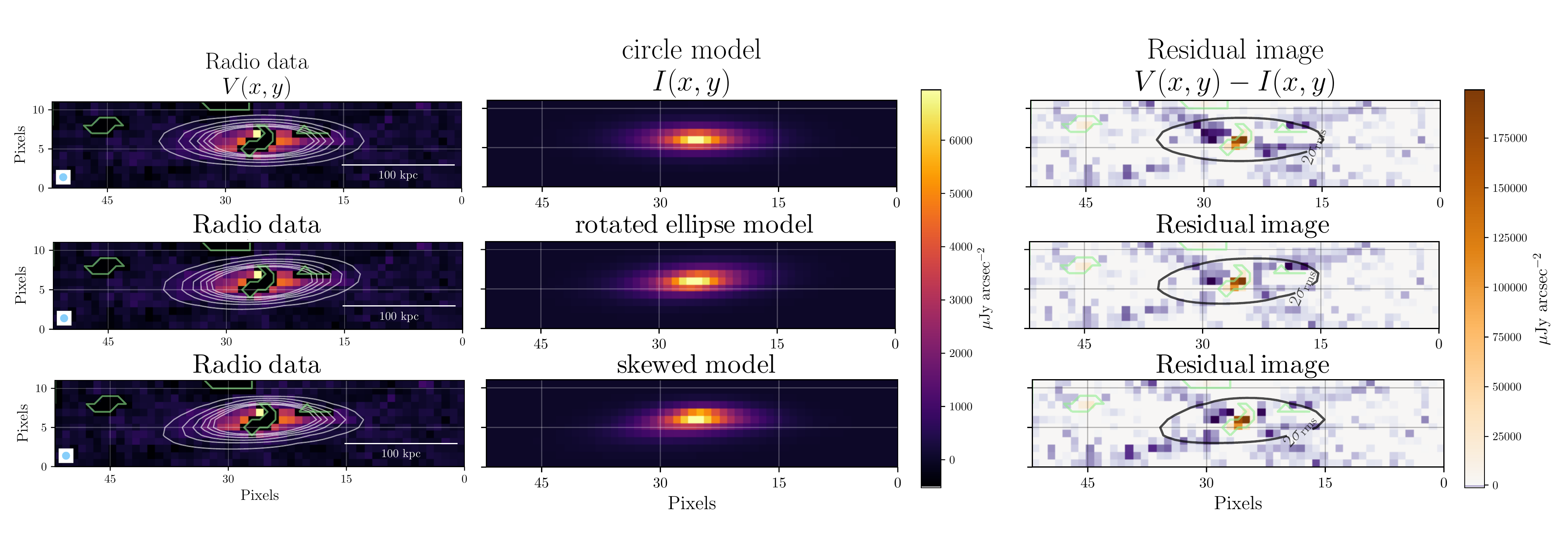}
	\vspace{-1.cm}
		\caption{\footnotesize Same as Fig. \ref{fig:abell_models}. Fitting result for the \emph{Phoenix cluster} at 1.5~GHz.}
	\label{fig:phoenix_models}
\end{figure*}

The used image has a strong contaminating source right at the centre of the mini-halo, which is carefully masked out as shown in figure~\ref{fig:phoenix_all}. The detailed results per fit are shown in Figure~\ref{fig:phoenix_models}. This figure highlights the effect an elongated beam has on regridding. 

{\tt Halo-FDCA} returns a flux density between $8.1-8.4$ mJy over the different models. Despite the small size of the mini-halo (there are only a few beams across the halo in one direction) and central contamination, runs for all models result in precise flux densities that are in agreement with each other. The $\chi^2_{\text{red}}$ values for both the circular and skewed model show values very close to one, while the elliptical model turns out slightly over-subtracted in the sense that the model is locally larger than the data (visualized in residual images). This resulted in a relatively high flux density for this model. As shown in Figure~\ref{fig:phoenix_models}, the skewed model appears to have the most over-subtraction within $2\sigma_{\text{rms}}$ of the diffuse emission. We therefore adopt an estimated flux density of $S_{\text{1.5 GHz}}=8.15\pm0.18$ mJy (statistical uncertainties only) with $r_e= \{39\pm2,\;21\pm2,\;28\pm1,\;20\pm1\}$ kpc. This agrees within the uncertainties with the value of $8.5\pm0.9$ reported by \citet{2020arXiv200913238T} which was obtained by fitting a Gaussian profile.

\subsection{The influence of masking}
\label{sec:mask_results}
In Section~\ref{sec:code} we discussed how to separate the radio halo emission from contaminating sources, either by masking extended sources or by using compact source subtracted images made with a uv-cut. 

\begin{figure*}[h]
	\includegraphics[width=.45\textwidth]{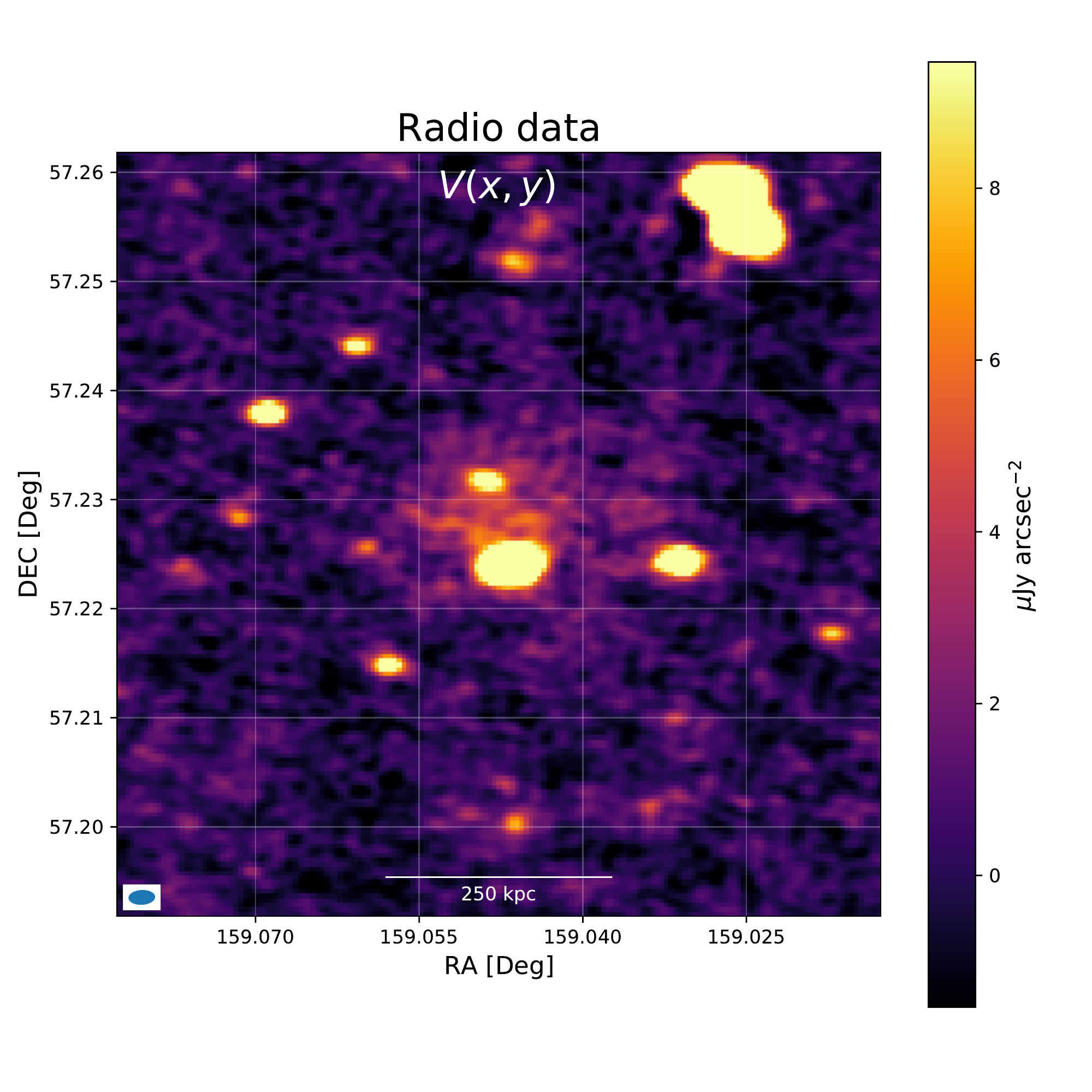}
	\includegraphics[width=.45\textwidth]{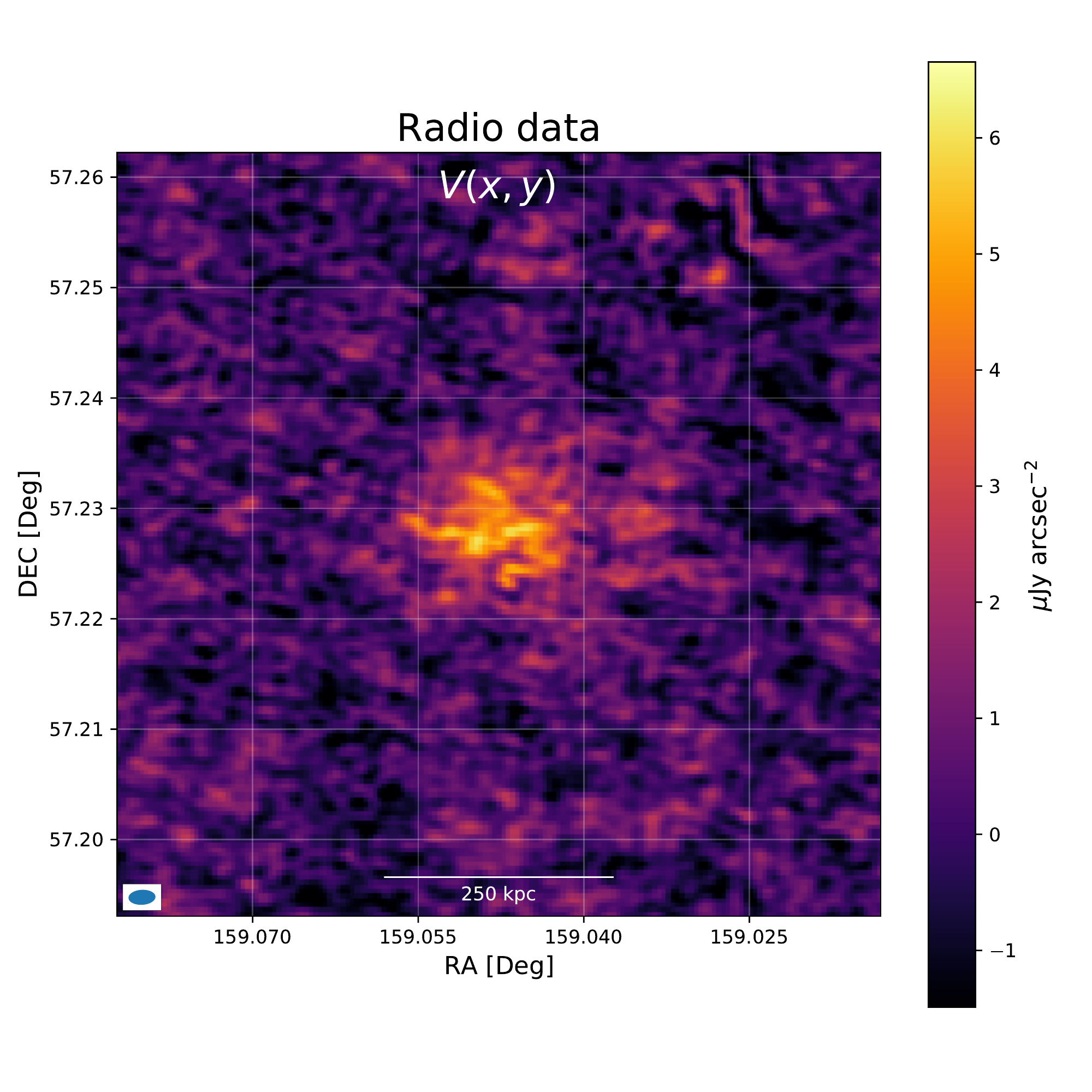}
		\caption{\footnotesize 144\,MHz LOFAR images of MCXC~J1036.1+5713 ($z=0.203$). \emph{Left panel:} Original data without uv-cut with noise $\sigma_{\text{rms}}=36.36$ $\mu$Jy beam$^{-1}$.   Bright compact sources can be seen embedded in the radio halo. \emph{Right panel:} 144\,MHz LOFAR images of MCXC~J1036.1+5713 with noise $\sigma_{\text{rms}}=37.43$ $\mu$Jy beam$^{-1}$. Compact sources were subtracted from the uv-data before making the image. The beam sizes are indicated at the bottom corners.}
	\label{fig:mcxcj1036_image}
\end{figure*}

\begin{figure*}[h]
	\includegraphics[width=1.\textwidth]{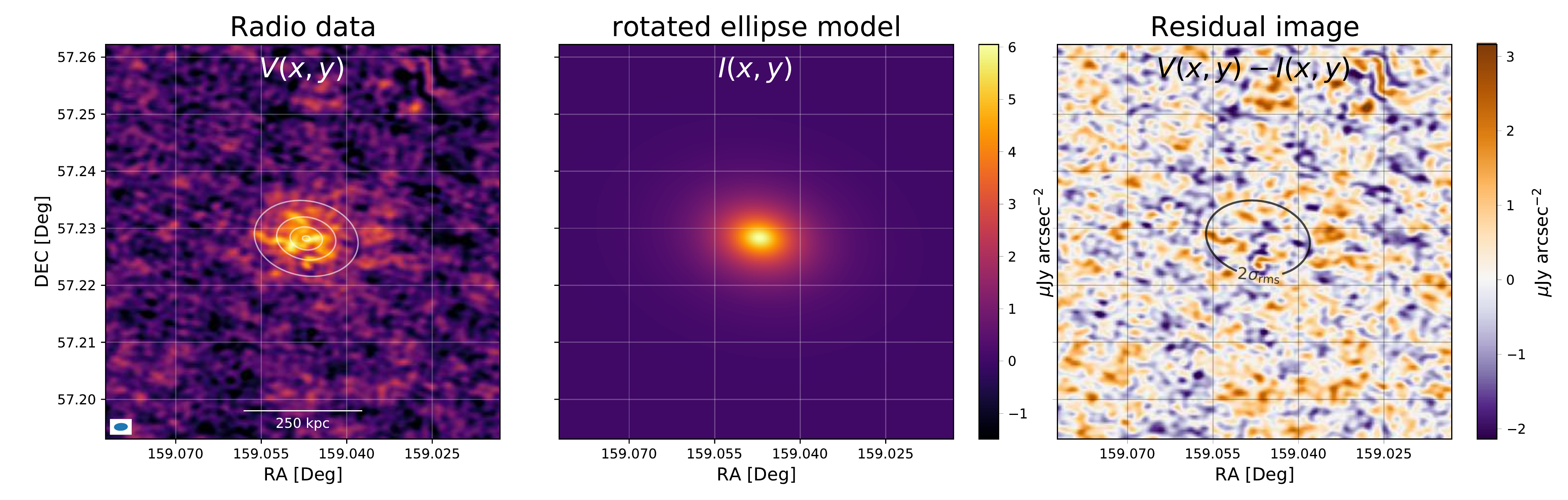}
	\includegraphics[width=1.\textwidth]{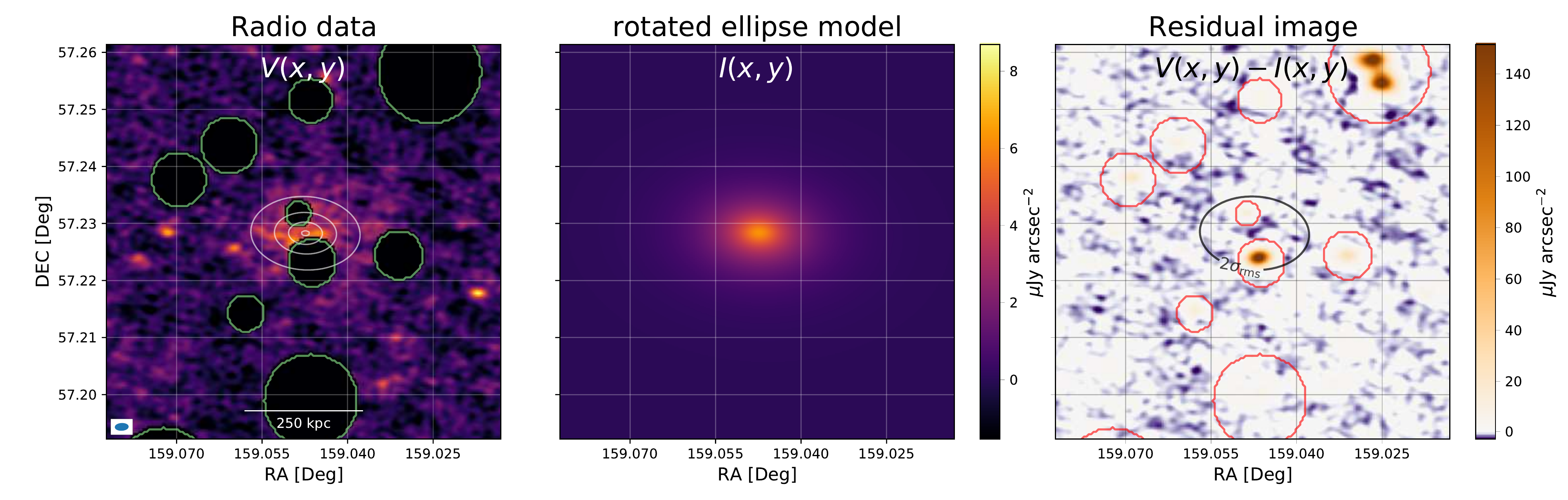}
	\includegraphics[width=1.\textwidth]{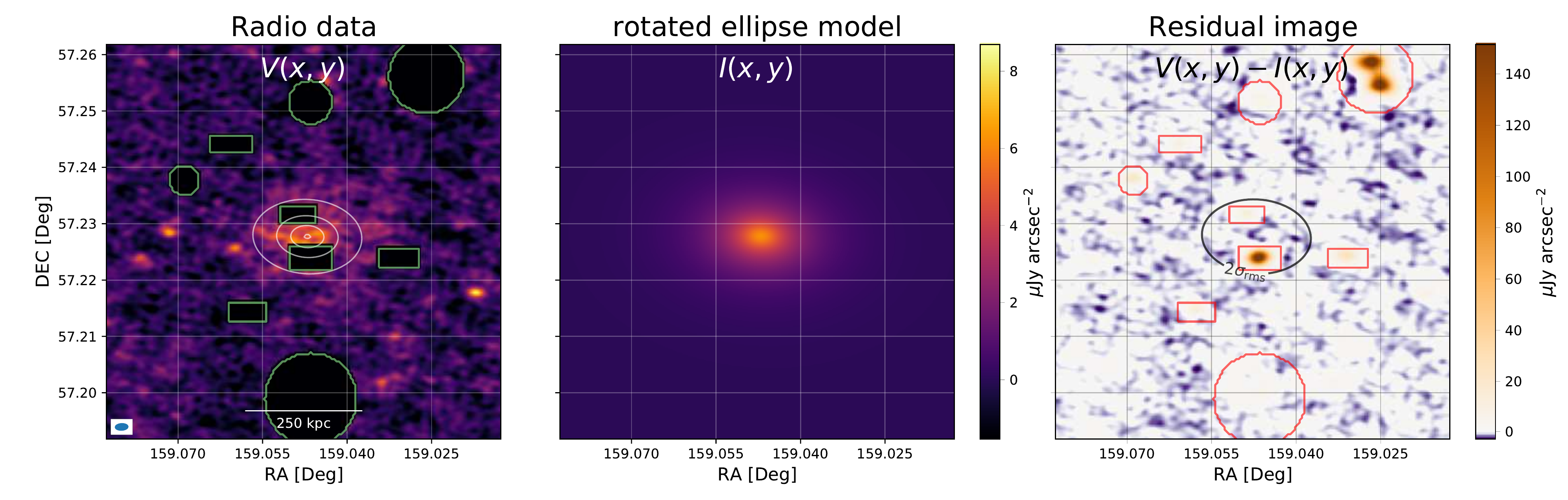}
		\caption{\footnotesize Results for MCXC~J1036.1+5713 using the two images from figure~\ref{fig:mcxcj1036_image}. All rows are similar to Figure~\ref{fig:Abell_all}. \emph{Top row:} Fit to the compact source subtracted data without any other masking. Relevant parameters are estimated to be $I_0= 7.47\pm0.35$ $\mu$Jy arcsec$^{-2}$, $r_x = 68\pm4$ kpc and $r_y = 49\pm3$ kpc. The flux density is $13.96\pm0.63$ mJy. \emph{Middle row:} Fit to the original data with relatively broad masking around the compact emission. Relevant parameters are estimated to be $I_0= 7.66\pm0.44$ $\mu$Jy arcsec$^{-2}$, $r_x = 71\pm4$ kpc and $r_y = 48\pm3$ kpc. The flux density is $14.76\pm0.75$ mJy. \emph{Bottom row:} Fit to the original data with relatively narrow masking. Relevant parameters are estimated to be $I_0= 7.55\pm0.43$ $\mu$Jy arcsec$^{-2}$, $r_x = 73\pm4$ kpc and $r_y = 50\pm3$ kpc. The flux density is $15.32\pm0.70$ mJy.}
	\label{fig:mcxcj1036_results}
\end{figure*}

Here, we perform a test to show how the different methods affect the fitting. For this we used 144\,MHz LOFAR data of MCXC\,J1036.1+5713 from \citet{2020arXiv201108249O}. The compact source subtracted image from this cluster is quite clean for fitting purposes (see Figure~\ref{fig:mcxcj1036_image}, right panel), while the unsubtracted data shows some compact sources embedded in the halo (see Figure~\ref{fig:mcxcj1036_image}, left panel). We estimated the flux density of the halo in both images using two different masks for the left panel of Figure~\ref{fig:mcxcj1036_image} to assess the influence of the adopted mask on the results. The fit results are shown in Figure~\ref{fig:mcxcj1036_results}. The compact source subtracted flux density is $13.96\pm0.63$ mJy, the flux density for the 'narrow' mask is $15.32\pm0.70$ mJy and the flux density for the 'broad' mask is $14.76\pm0.75$ mJy. This shows that the flux density is not significantly influenced by the specific mask.

\section{Discussion}
\label{sec:discussion}
Along with the Bayesian methodology, the {\tt Halo-FDCA} code introduces more generalized equations to describe the surface brightness of radio halos in galaxy clusters than the model proposed by \cite{radial_profile}. This generalization included the introduction of skewed asymmetric models and the form factor $k$. For the sake of comparison, this extra parameter was not used in the examples. The inclusion of this parameter option could prove to be useful in cases the user desires more model shape flexibility. This option could also be used with a sample of halos to study the distribution of $k$. 

When estimating the flux densities, we have made extensive use of masks, especially for the Phoenix cluster. While the pipeline itself mostly functions without user-based biases, masking can introduce such biases as they are arbitrarily drawn. Although we did not extensively study the influence of masking on the estimation of the flux density, we expect that in most cases the masking does not significantly influence the results based on the test discussed in Section~\ref{sec:mask_results}.
In cases like MCXC1036.1+5713 and the Phoenix cluster, where contamination is present at the halo centre, it is advised to make sure all contamination is masked out. The code can extrapolate the model where data is missing but will falsely include contamination if it is not masked out properly. It is better to mask out too much than too little. The uncertainty on the flux density takes into account the level of masking adopted, becoming larger when larger areas are masked.

The choice for a standardized integration radius is to be made by the user. Although infinity is a natural choice from a mathematical point of view, it is not necessarily the best choice if accuracy is desired and the aim is to compare different radio halos. In particular, for radio halos detected at low signal to noise, the extrapolation to infinity might introduce unnecessary large uncertainties since most flux density comes from parts of the radio halo that are below the noise level of the map.  Beyond a certain radius from the center of the halo, the model is essentially not constrained by the data in this case. Therefore, it might be better to intentionally integrate up to distance $d$ since this value relies less on extrapolation. For example, \cite{radial_profile} adopted $3r_{e}$ where the surface brightness is 5\% of its peak value (for the simple exponential model). This might be even too far out for barely detected halos. The flexible code allows the user to adopt their own choice of integration radius in units of $r_{e}$.

The merging of two clusters, each with its own radio halo might introduce a situation where two radio halos overlap and joint fitting is necessary. Currently, {\tt Halo-FDCA} is not able to fit more than one radio halo ``component'' but if required the code can be expanded to provide such functionality, keeping the overall framework intact. This would also allow the fitting of a smaller mini-halo component embedded in a larger scale diffuse component, possible examples of such a case are the clusters Abell\,2142 \citep[][]{2017AA...603A.125V} and PSZ1G139.61+24.20 \citep{2018MNRAS.478.2234S}. Additionally, the overlap of a radio ``bridge'' \citep[e.g.,][]{2018MNRAS.478..885B,2020MNRAS.tmpL.159B,2019Sci...364..981G} with a radio halo might require adding another model component. However, determining what the proper functional form is for fitting a radio bridge requires more investigation.

\section{Conclusions}
\label{sec:conclusion}
We presented {\tt Halo-FDCA}, a fitting algorithm based on Bayesian inference to accurately estimate the flux density of radio halos in galaxy clusters. Bayesian inference methods adopted here find parameters that best fit the data based on the conditional probability of observed data given the model. The implementation used here makes use of Markov Chain Monte Carlo fitting. %

A total of three different models are included for fitting. These models include circular, elliptical and skewed exponential distributions. Halos are detected and identified by likelihood estimation rather than subjective judgment, which should allow for a better comparison of radio images from the same objects but at different resolutions and depths. The algorithm also includes a flexible option to mask any desired portion of the image. 

The presented models for flux density estimation can be of great importance for deriving accurate e statistical properties of large samples of radio halos which should help us to better understand the origin of these enigmatic sources.
\newline

{\footnotesize 
\emph{Acknowledgements}
R.J.v.W. acknowledges support from the ERC Starting Grant ClusterWeb 804208. A.B. and R.J.v.W. acknowledge support from the VIDI research programme with project number 639.042.729, which is financed by the Netherlands Organisation for Scientific Research (NWO). We thank Erik Osinga for testing and feedback on the code. This research made use of \texttt{Astropy},\footnote{http://www.astropy.org} a community-developed core \texttt{Python} package for Astronomy \citep{2013A&A...558A..33A,2018AJ....156..123A} and \texttt{Scipy},\footnote{https://scipy.org/index.html} which has fundamental algorithms for scientific computing in \texttt{Python} \citep{2020SciPy-NMeth}.
}

\printcredits

\bibliographystyle{cas-model2-names}

\onecolumn

\appendix{
\renewcommand{\thefigure}{\thesection.\arabic{figure}}
\setcounter{figure}{0}    
\section{Exponential Profiles and Their Integrals}
\label{app:exponential_functions}

The total flux density of $I(x,y)$ is written as

\begin{equation}
	S_{\nu} = \int\int I(x,y)\;dx\;dy.
\end{equation}

The skewed model is integrated here to derive a general expression for the total flux density. For this calculation, $\phi$, $x_0$ and $y_0$ are set to zero since those parameters only transpose and rotate the distribution and do not actually change its shape. This simplifies Eq. \eqref{8D-model} significantly, reducing the function $G$ to 

\begin{equation}
  G(\vec{r})^{1/(0.5+k)}=\begin{cases}
    \left(\dfrac{x}{r_{x^{+}}}\right)^2+\left(\dfrac{y}{r_{y^{+}}}\right)^2, & \text{if $X$, $Y\geq0$}.\\
    
    \left(\dfrac{x}{r_{x^{-}}}\right)^2+\left(\dfrac{y}{r_{y^{+}}}\right)^2, & \text{if $X\leq0$, $Y>0$}.\\
    
    \left(\dfrac{x}{r_{x^{-}}}\right)^2+\left(\dfrac{y}{r_{y^{-}}}\right)^2, & \text{if $X$, $Y<0$}.\\
    
    \left(\dfrac{x}{r_{x^{+}}}\right)^2+\left(\dfrac{y}{r_{y^{-}}}\right)^2, & \text{if $X>0$, $Y\leq0$}.
  \end{cases}
\label{8D-model}
\end{equation}

To make matters simpler, an ellipse with major and minor axis corresponding to the $e$-folding radii of a certain quadrant are integrated over that quadrant only. This results in the evaluation of four integrals which all integrate over the elliptic exponential equation in a specific quadrant. This looks like

\begin{equation}
	S_{\nu} = 
		\int^{\infty}_{0}\int^{\infty}_{0}I_1(x,y)\;dx\;dy
		\int^{0}_{-\infty}\int^{\infty}_{0}I_2(x,y)\;dx\;dy
		\int^{\infty}_{0}\int^{0}_{-\infty}I_3(x,y)\;dx\;dy
		\int^{0}_{-\infty}\int^{0}_{-\infty}I_4(x,y)\;dx\;dy
\end{equation}
Where $I$ corresponds to the elliptical model. It is important to note that every function $I_i$ that is appearing, has different $e$-folding radii. Formally, limits have to be taken to let the functions approach zero, That kind of notation is omitted here because, in the end, it will give the same result as  when it would be included.

Another step to simplify the integrals themselves is to note that the elliptic function is symmetric in every quadrant and has the exact same shape in all four of them. Using this the integrals can be taken over all the real numbers again with a factor of $1/4$ in front to account for the fact that only one quadrant is taken into account. This results in  

\begin{equation}
	S_{\nu} = \frac{1}{4}\int^{\infty}_{-\infty}\int^{\infty}_{-\infty}I_1(x,y)\;dx\;dy+
	\frac{1}{4}\int^{\infty}_{-\infty}\int^{\infty}_{-\infty}I_2(x,y)\;dx\;dy+
	\frac{1}{4}\int^{\infty}_{-\infty}\int^{\infty}_{-\infty}I_3(x,y)\;dx\;dy+
	\frac{1}{4}\int^{\infty}_{-\infty}\int^{\infty}_{-\infty}I_4(x,y)\;dx\;dy
\end{equation}

We continue to evaluate one integral $S_i$ with function $I_i$ which has $e$-folding radii $r_x$ and $r_y$. Writing out $I(x,y)$ yields

\begin{align}
	S_i = \frac{I_0}{4}\int^{\infty}_{-\infty}\int^{\infty}_{-\infty}\exp\left[-\left(\frac{x^2}{r^2_x}+\frac{y^2}{r^2_y}\right)^{0.5+k}\right]\;dx\;dy
\end{align}

We transform to elliptical polar coordinates ($\rho$, $\theta$) where we substitute $x=\rho r_x\cos{\theta}$ and $y=\rho r_y\sin{\theta}$. It can be checked that the Jacobian for this coordinate transformation is $r_x r_y\rho$ such that $dx\;dy=r_x r_y\rho\;d\rho\;d\theta$. This is the point where the radius of integration $d$ can be specified, since $\rho$ is a radial coordinate. This $d$ value can in theory be different for every quadrant, but it is for this application highly advised to stick to one value for $d$ for every integral. The integral now becomes

\begin{align}
	S_i &= \frac{I_0}{4}\int^{2\pi}_{0} \int^{d}_{0}\exp\left[-\left(
	\frac{\rho^2 r^2_x\cos^2{\theta}}{r^2_x}+
	\frac{\rho^2 r^2_y\sin^2{\theta}}{r^2_y}\right)^{0.5+k}\right]r_x r_y\rho\;d\rho\;d\theta\\
	&=\frac{I_0r_x r_y}{4}\int^{2\pi}_{0} \int^{d}_{0}\exp\left[-\left(
	\rho^2 \cos^2{\theta}+
	\rho^2 \sin^2{\theta}\right)^{0.5+k}\right]\rho\;d\rho\;d\theta\\	
	&=\frac{I_0r_x r_y}{4}\int^{2\pi}_{0}d\theta \int^{d}_{0}\exp\left[-\left(\rho^2\right)^{0.5+k}\right]\rho\;d\rho\\
	&=\frac{I_0\pi r_x r_y}{2} \int^{d}_{0}\rho\exp\left(-\rho^{m}\right)\;d\rho
\end{align}
 where in the last line, $m\equiv1+2k$ \footnote{For $k=0$, the integral is simply solved as it reduces to $\int_0^d \rho e^{-\rho} dx = 1 - (d + 1) e^{-d}$ which equals one as $d\rightarrow\infty$. \label{footnote:appendix}}.
The resulting integral is a standard one that can be evaluated fairly easily using $u=\rho^m$ as a substitution. After substitution, the integral becomes 

\begin{align}
	S_i &=\frac{I_0\pi r_x r_y}{2m} \int^{d^m}_{0} u^{2/m-1}e^{-u}\;du.
\end{align}
This integral is generally known as the lower incomplete gamma function $\gamma(x,k)$, defined as
\begin{equation}
    \gamma(x,k)\equiv\int_0^k t^{x-1}e^{-x}\;dt.
\end{equation}
The expression for one of the integrals now becomes
\begin{equation}
	S_i = \gamma\left(\frac2m,d^m\right)\frac{I_0\pi}{2m}r_xr_y.
\end{equation}
Now, we only have to add the four integrals with different $e$-folding radii and equal $d$ together to obtain the final expression:
\begin{equation}
	S_{\nu}(d) = \gamma\left(\frac2m,d^m\right)\frac{I_0\pi}{2m}(r_{x^+}r_{y^+}+r_{x^-}r_{y^+}+r_{x^+}r_{y^-}+r_{x^-}r_{y^-})
\end{equation}

\setcounter{figure}{0}    
\newpage
\section{Figures Accompanying Results}
\label{app:results}

 \begin{figure*}[h]
	\includegraphics[width=1.\columnwidth]{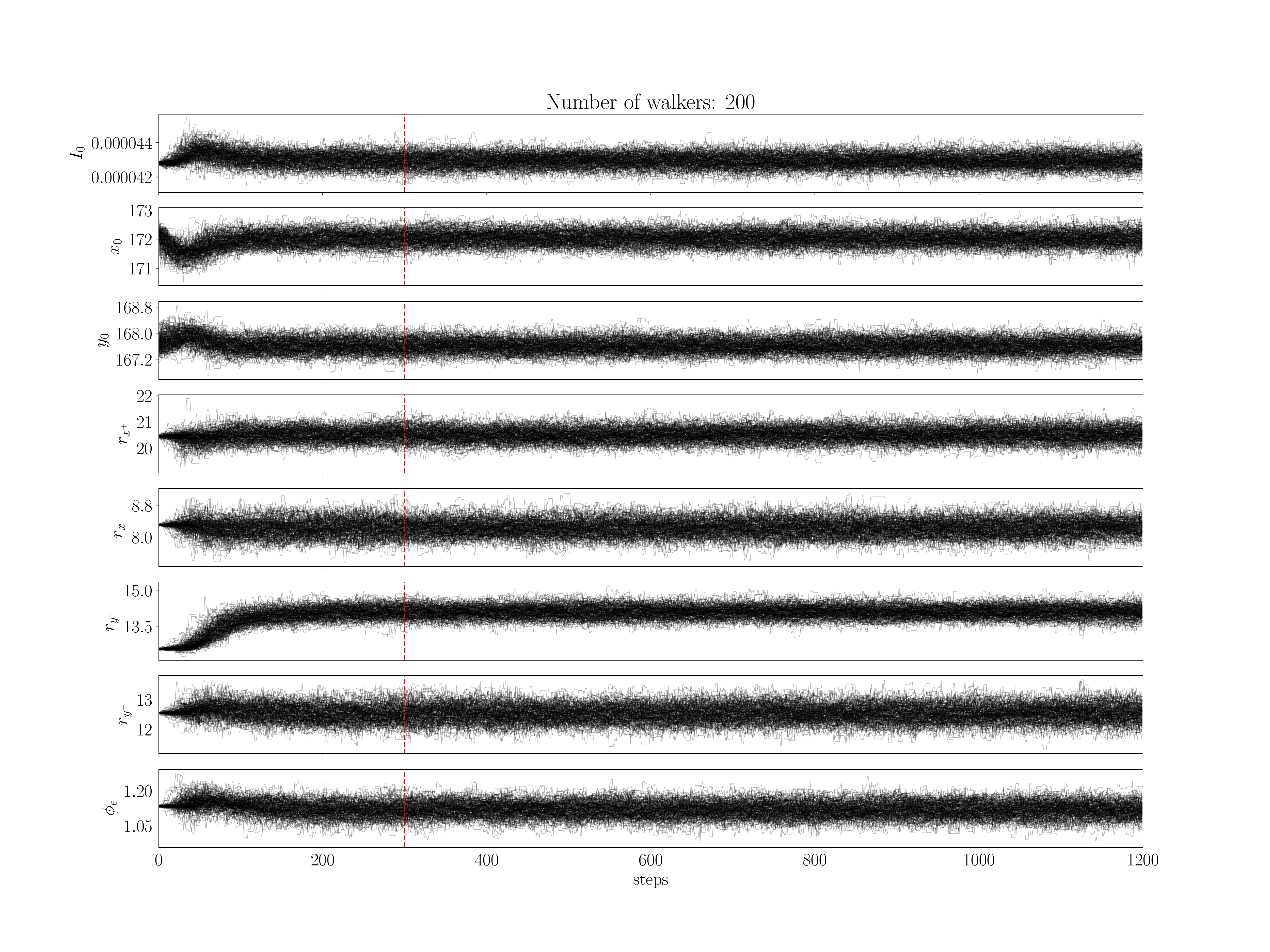}
		\caption{\footnotesize Example of a walker plot which is a direct result from a MCMC fitting with 200 walkers and 1200 evaluations. This particular example shows walker movements while fitting an elliptical model to Abell\,2744. It can be seen in panel 2 and 6 for instance, that it takes some time before the algorithm has settled on a certain value. This is after around 200 steps. The default burn-in setting is a quarter of the total number of steps (red line), thus only values evaluated after step n/4 are used in estimating the parameters.}
	\label{fig:walker}
\end{figure*}

 \begin{figure*}[h]
	\includegraphics[width=1\columnwidth]{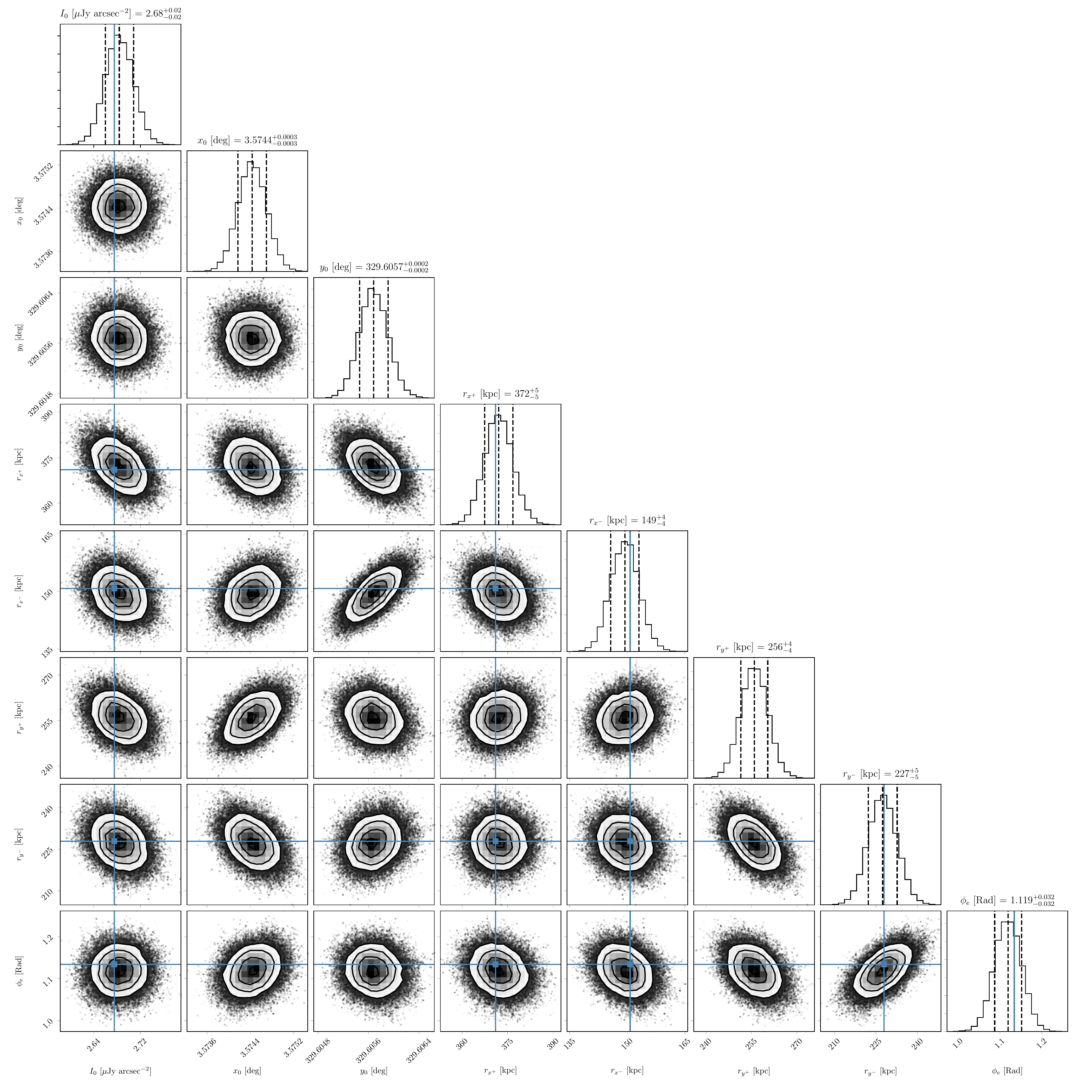}
  	\vspace{-0.4cm}
		\caption{\footnotesize Example of a corner plot which is a direct result from a MCMC fitting with 200 walkers and 1200 evaluations. This particular example shows walker movements while fitting a skewed model to Abell\,2744. Blue lines indicate initial guesses resulting from the pre-MCMC fit that are passed to \texttt{emcee}. Histograms showing the distribution of maximum likelihood parameters. The two-dimensional scatter plots show how every parameter solution is correlated. %
		The solutions of $I_0-r_x$ and $I_0-r_y$ are in fact slightly correlated as illustrated by the elongated scatter distribution. This figure shows that a pre-MCMC fit does not always provide the most accurate initial guess (in this case for $x_0$, $y_0$ and $r_{y^+}$). The code was nevertheless able to diverge from the initial guess and find a more likely model. %
		}
	\label{fig:corner}
\end{figure*}}

\end{document}